\documentclass[11pt]{elsarticle}
\usepackage{epsfig,amsfonts,amssymb,amsmath}
\usepackage{color}
\usepackage{array}
\usepackage{lineno}
\usepackage{multirow}

\newtheorem{theorem}{Theorem}[section]

\newtheorem{lemma}[theorem]{Lemma}
\newtheorem{corollary}[theorem]{Corollary}

\makeatletter
\def\ps@pprintTitle{%
  \let\@oddhead\@empty
  \let\@evenhead\@empty
  \let\@oddfoot\@empty
  \let\@evenfoot\@oddfoot
}
\makeatother

\begin{document}

\begin{frontmatter}

\title{Analysis of an epidemic model with awareness decay \\ on regular random networks}

\author[address1]{David Juher}
\ead{david.juher@udg.edu} 

\author[address2]{Istvan Z. Kiss}
\ead{I.Z.Kiss@sussex.ac.uk}

\author[address1]{Joan Salda\~{n}a\corref{cor1}}
\ead{joan.saldana@udg.edu}
\cortext[cor1]{Corresponding author}

\address[address1]{Departament d'Inform\`{a}tica, Matem\`{a}tica Aplicada i Estad\'istica
\\
Universitat de Girona, Catalonia, Spain}

\address[address2]{School of Mathematical and Physical Sciences, Department of Mathematics
\\
University of Sussex, Falmer, Brighton BN1 9QH, UK}

\begin{abstract}
The existence of a die-out threshold (different from the classic disease-invasion one) defining a region of slow extinction of an epidemic has been proved elsewhere for susceptible-aware-infectious-susceptible models without awareness decay, through bifurcation analysis. By means of an equivalent mean-field model defined on regular random networks, we interpret the dynamics of the system in this region and prove that the existence of bifurcation for this second epidemic threshold crucially depends on the absence of awareness decay. We show that the continuum of equilibria that characterizes the slow die-out dynamics collapses into a unique equilibrium when a constant rate of awareness decay is assumed, no matter how small, and that the resulting bifurcation from the disease-free equilibrium is equivalent to that of standard epidemic models. We illustrate these findings with continuous-time stochastic simulations on regular random networks with different degrees. Finally, the behaviour of solutions with and without decay in awareness is compared around the second epidemic threshold for a small rate of awareness decay. 
\end{abstract}


\end{frontmatter}



\section{Introduction}\label{s1}

The study of the impact of behavioural responses on the progression of infectious diseases in human populations has received a lot of interest during the last years. An important challenge is how to capture features of human behaviour in epidemic modelling \cite{Ferguson}. In a broad sense, when dealing with sexually transmitted diseases, a high heterogeneity in the number of contacts reflects the high variability of individuals' behaviour  \cite{Anderson}. But, in addition to this relevant aspect, individual responses adopted to reduce the perceived risk of contagion constitute another facet of human behaviour that greatly impacts the dynamics of an epidemic. Social avoidance behaviours, for instance, were prevalent in Hong Kong during the SARS pandemic and, also, at the initial stage of the H1N1 epidemic \cite{Lau}. 

Avoidance of contacts with infectious individuals is another example of behavioural plasticity. This social distancing leads to the idea of link rewiring and is one of the basis of adaptive networks in which the structure of the contact pattern evolves with the progression of the epidemic \cite{Gross06, Llensa}. Disease-avoiding rewiring implicitly assumes the local knowledge and transmission of information which allows individuals to assess the status of their nearest neighbours. Another way to capture information transmission is by accounting for the difference in how information about the disease is processed and acted upon. For example, it is reasonable to assume that risk perception is heterogeneous across a population, meaning that those who are more risk averse are more likely to adopt preventive measures against contagion. Such individuals have been labelled as \textit{alert} or \textit{aware} in the specialist literature.  Some models consider that both susceptible and infectious individuals can be in either of two states: aware/responsive and unaware/non-responsive \cite{Funk09,Funk10,Granell,Kiss10}. If information dissemination is explicitly taken into account, new layers are added to the basic contact network. This extension results in dealing with overlapping or multiplex networks where the disease and information dissemination networks overlap to different levels \cite{Granell, Sahneh14, Sahneh12b}.

Predictions from these approaches vary and depend on the particular modelling assumptions. For instance, in \cite{Funk09} the authors assume a decreasing quality of the information and fading of awareness and claim that, below a critical infection rate, awareness and a lower susceptibility of aware individuals lead to a reduction in the basic reproduction number $R_0$ of a susceptible-infectious-recovered (SIR) model. Interestingly, this claim follows under an individual-based approach but not from a mean-field analysis. A similar conclusion is obtained in \cite{Kiss10} from a mean-field model which assumes that both susceptible and infectious individuals can be either responsive or non-responsive against the disease. In this setting, a contact-based transmission of information can change, under suitable conditions, the epidemic threshold and prevent the spread of the disease.

The approach that motivates our study is the one introduced in \cite{Sahneh12a, Sahneh11}. In these papers, the authors assume a new class of individuals, the aware or alerted ones, which are non-infectious individuals with a reduced susceptibility. Awareness arises in susceptible individuals when they have infectious neighbours, that is, from a contact-based transmission of information. Aware individuals can get infected at a lower infection rate compared to fully susceptible ones, due to the adoption of preventive measures.  Formulating the network epidemic model in terms of an approximation of the exact Markov processes, the authors prove the existence of a second threshold (different from the classic disease-invasion one) for the preventive response to suppress epidemic spreading.  Certainly, this is a result that has not been observed in previous modelling approaches and seems to give an important role to aware people at the beginning of the epidemic.

The aim of this paper is to give a simple but full interpretation of the existence of such an epidemic threshold using a mean-field approach to the original equations. In contrast to previous results (see, for instance, \cite{Funk09}), in Sections \ref{model} and \ref{SAIS-RN} we show that the mean-field model captures the dynamics of the whole system and leads to the same epidemic thresholds as the network model for fully connected and regular random networks. Therefore, it offers a simple and complete description of the epidemic dynamics (Section \ref{SAIS-no-decay}). In particular, this mean-field formulation allows us to prove by means of Peixoto's theorem (see \cite{Perko}) that the original model introduced in \cite{Sahneh12a,Sahneh11} defined on regular random networks is structurally unstable because it has a continuum of equilibria and, so, the qualitative behaviour of its solutions can change with a small and smooth perturbation of its equations. In Section \ref{SAIS-with-decay}, we show that including an awareness decay leads to the disappearance of the second epidemic threshold, and to the appearance of oscillations around an endemic equilibrium instead of the occurrence of minor outbreaks associated to an slow die-out of the epidemic \cite{Sahneh12a}. The analysis of many real outbreak episodes, such as the SARS epidemic in Hong-Kong in 2003 and HIV/AIDS, provide clear evidence of awareness decay over time. Namely, for the former, fewer people are wearing face masks when they have a cold/flu and washing their hands regularly. While for the latter, there are still a large number of individuals adopting risky behaviours despite the heightened level of awareness and prevention campaigns \cite{Erinosho}. Thus, the inclusion of awareness decay is an important model ingredient and needs to be accounted for.

\section{The SAIS model} \label{model}

We consider a population of size $N$ for which an individual $i$ can be in one of the following states: $S_i$, $I_i$, or $A_i$, where capital letters denote the \emph{susceptible}, \emph{infectious}, or \emph{aware/alert} states, respectively. We also consider the following parameters
\begin{itemize}
\item per-contact infection rate for susceptible individuals, $\beta_0 > 0$,
\item per-contact infection rate for aware individuals, $\beta^0_a > 0$,
\item per-contact awareness rate for susceptible individuals, $\kappa_0 > 0$,
\item recovery rate for infectious individuals, $\delta > 0$,
\item rate of awareness decay for aware individuals, $\delta_a \ge 0$,
\end{itemize}
together with the following rules for the epidemic/awareness evolution:
\begin{equation}\label{rules}
I_i \stackrel{\delta}{\longrightarrow} S_i,\,\, A_i \stackrel{\delta_a}{\longrightarrow} S_i,\,\, I_i+S_j \stackrel{\beta_0}{\longrightarrow} I_i + I_j,\,\,
I_i + A_j \stackrel{\beta^0_a}{\longrightarrow} I_i + I_j,\,\, I_i + S_j \stackrel{\kappa_0}{\longrightarrow} I_i + A_j.
\end{equation}
Finally, to reflect the fact that aware individuals get infected at a lower rate than unaware (susceptible) ones, we will assume  $\beta^0_a < \beta_0$ when necessary. From a mathematical point of view, this hypothesis restricts the possible behaviours of the solutions of the model and will prevent us from results which do not make biological sense. An example of these results happens for $\delta_a > 0$ if $\beta_a > \delta$ because, then, it is possible to have a stable endemic equilibrium which is sustained thanks to a higher susceptibility ($\beta^0_a > \beta_0$) of aware individuals.     

If $p_{S_i}$, $p_{I_i}$, $p_{A_i}$ are the probabilities for node $i$ to be susceptible, infectious, and aware, respectively, $p_{S_iI_j}$ is the joint probability for node $i$ being susceptible and for a neighbour $j$ being infectious (and similarly for $p_{A_iI_j}$), and $(a_{ij})_{i,j=1,\dots,N}$ is the adjacency matrix of the contact network, the exact model for the setup above in a continuous-time setting is given by:
\begin{eqnarray*}
\displaystyle
\frac{d p_{S_i}(t)}{dt}& = & - \sum_{j=1}^N  \beta_{ij} \, a_{ij} p_{S_iI_j}(t) - \sum_{j=1}^N \kappa_{ij} \, a_{ij} p_{S_iI_j}(t)  + \delta_{i} \, p_{I_i}(t) + \delta_{i}^{a}  \, p_{A_i}(t) \, ,\\ \\
\displaystyle
\frac{d p_{I_i}(t)}{dt} & = & \sum_{j=1}^N  \beta_{ij} \, a_{ij} p_{S_iI_j}(t) + \sum_{j=1}^N \beta_{ij}^{a} \, a_{ij} p_{A_iI_j}(t) - \delta_i  p_{I_i}(t) \, ,\\ \\
\displaystyle
\frac{d p_{A_i}(t)}{dt} & = &  \sum_{j=1}^N \kappa_{ij} \, a_{ij} p_{S_iI_j}(t)  - \sum_{j=1}^N  \beta_{ij}^{a} \, a_{ij} p_{A_iI_j}(t) - \delta_i^a p_{A_i}(t) \, ,
\end{eqnarray*}
where, for sake of generality in the presentation, the transmission rates $\beta$, $\beta_a$, and $\kappa$ are assumed to be dependent on the involved pair $(i,j)$ of individuals. In what follows, we ignore $p_{S_i}(t)$, since $(p_{S_i}+p_{I_i}+p_{A_i})(t)=1$ for $i=1,2,\dots,N$. This is not a closed or self-consistent system as further equations for the pairs are needed. To avoid this dependence on higher order moments, let us assume that the joint probability can be written as $p_{S_iI_j} = p_{S_i} \cdot p_{I_j}$ (or $p_{A_iI_j} = p_{A_i} \cdot p_{I_j}$), that is, it is independent of the neighbourhood configuration of node $i$ and node $j$.

Assuming the same rates $\beta_0$, $\kappa_0$, and $\beta^0_a$ of transmission across a contact and the same recovery and awareness decay rates, $\delta$ and $\delta_a$, for all the nodes, denoting $p_{I_i}=p_i$ and $p_{A_i} = q_i$, and dropping the time dependence for simplicity, the model reads:
\begin{equation}\label{markov}
\left\{\begin{array}{lll}
\displaystyle
\frac{dp_i}{dt} & = & \beta_0 (1-p_i-q_i) \sum_j a_{ij} p_j + \beta^0_a q_i \sum_j a_{ij}p_j - \delta p_i \, ,
\\ \\
\displaystyle
\frac{dq_i}{dt} & = & \kappa_0 (1-p_i-q_i) \sum_j a_{ij} p_j - \beta^0_a q_i \sum_j a_{ij}p_j - \delta_a q_i \, .
\end{array}\right.
\end{equation}
for $i=1,\dots,N$. For $\delta_a = 0$, these equations define the so-called SAIS epidemic model considered in \cite{Sahneh12a, Sahneh11}.


\section{The SAIS model on regular random networks} \label{SAIS-RN}

In order to give a simple interpretation of the the epidemic thresholds obtained in \cite{Sahneh12a,Sahneh11}, from now on we will restrict ourselves to regular random networks, that is, networks where all the nodes have the same number of neighbours (degree), $k$, which are randomly chosen among the nodes in the network. Our approach will also embrace fully connected networks  ($k=N-1$). In this paper, we will not focus on the goodness of the mean-field approximation $p_{S_iI_j} = p_{S_i} \cdot p_{I_j}$ and $p_{A_iI_j} = p_{A_i} \cdot p_{I_j}$, i.e. we will not investigate or quantify the error introduced by this closure. It is well know that such closures for particular dynamics and network topologies give a good approximation meaning that the exact model, via a Gillespie simulation, is well approximated by the closed model \cite{Givan}.

Since the degree is the only feature that characterizes a node, in this type of networks there is no reason to assume that some nodes have higher vulnerabilities than others. This means that it is reasonable to assume that any node of the network can get infected with the same initial probability $p_0$, and that the probability of being initially aware is also the same for all the nodes, namely, $q_0$. Under this uniform initial condition for $p_i$ and $q_i$, the following lemma states that these probabilities vary with time but are the same for any node in the network. 

\begin{lemma}
\label{equiv}
Consider the initial value problem (IVP) given by system \eqref{markov} defined on regular random networks and endowed with the initial condition $p_i(0)=p_0 \ge 0$ and $q_i(0)=q_0 \ge 0$  for $i = 1, 2,\dots, N$, and such that $p_0+q_0 \le 1$. The solution of this IVP is given by $(p_i(t), q_i(t)) = (p(t), q(t))$ $\forall \,i$ with $(p(t), q(t))$ being the solution of the system
\begin{equation}\label{markov2}
\left\{\begin{array}{lll}
\displaystyle
\frac{dp}{dt} & = & k \beta_0 (1-p-q) p + k \beta^0_a p \, q - \delta p \, ,
\\ \\
\displaystyle
\frac{dq}{dt} & = & k \kappa_0 (1-p-q) p - k \beta^0_a p \, q - \delta_a q \, ,
\end{array}\right.
\end{equation}
endowed with the initial condition $(p(0), q(0))=(p_0,q_0)$.
\end{lemma}

\textit{Proof}.
It is clear that if $(p_i(t), q_i(t)) = (p(t),q(t))$ $\forall \,i$ then $\sum_j a_{ij} p_j(t) = kp(t)$ $\forall \,i$ because each node has exactly $k$ neighbours, each one of them being infectious with probability $p(t)$. After introducing the value of this sum into system \eqref{markov}, it follows that $(p(t), q(t))$ must satisfy \eqref{markov2} with $(p(0), q(0))=(p_0,q_0)$ to be a solution to the IVP.

On the other hand, the standard theory of ODEs guarantees the existence and uniqueness of a local solution of the IVP defined by system \eqref{markov}, endowed with a non-negative initial condition $(p_i(0), q_i(0))=(p_0, q_0)$ $\forall i$, since the right-hand side (rhs) of system \eqref{markov} is quadratic in $p_i$ and $q_i$ and, hence, it is locally Lipschitz. Moreover, for $i=1,\dots,N$ and $(p_i,q_i) \in [0,1]\times [0,1]$, it immediately follows that $(dp_i/dt) |_{p_i=0} \ge 0$, $(dq_i/dt) |_{q_i=0} \ge 0$.  Finally, adding the equations of system \eqref{markov} for each $i$, we get $(d(p_i+q_i)/dt) |_{p_i+q_i = 1} < 0$. The region $\Omega := \{(p_1,q_1) \times \dots \times (p_N,q_N) \in [0,1]^{2n} | \, 0 \le p_i+q_i \le 1 \, \forall i\}$ is then positively invariant, which guarantees that the local solution can be extended to any $t>0$ for any initial condition in $\Omega$. Therefore, $(p_i(t), q_i(t)) = (p(t),q(t))$ $\forall \,i$ with $(p(t), q(t))$ satisfying \eqref{markov2} with $(p(0), q(0))=(p_0,q_0)$ turns out to be the unique global solution of the IVP.
$\Box$

From this lemma it follows that, under a uniform initial condition, the average number of infectious and aware individuals at time $t$ is given by $I(t)=\sum_i p_i(t)=N p(t)$ and $A(t)=\sum_i q_i(t)=N q(t)$. Then, we can consider the (expected) fractions of susceptible and aware individuals, $s=(N-I-A)/N$ and $a=A/N$, as state variables instead of working with the nodal probabilities of being infectious and aware. Indeed, $s(t)=1-p(t)-q(t)$ and $a(t)=q(t)$. In terms of $s(t)$ and $a(t)$, system \eqref{markov2} can be rewritten  as:
\begin{equation}\label{edos}
\left\{\begin{array}{lll}
\displaystyle \frac{ds}{dt} & = & (1-s-a)(\delta-(\kappa+\beta)s)+\delta_aa \\
& & \\
\displaystyle \frac{da}{dt} & = & (1-s-a)(\kappa s-\beta_aa)-\delta_aa,
\end{array}\right.
\end{equation}
where, as usual in mean-field models, $\beta = k \beta_0$, $\kappa = k \kappa_0$, and $\beta_a = k \beta^0_a$ are transmission rates per node and not per contact.

\begin{figure}[ht]
\centering
	\includegraphics[scale=0.4]{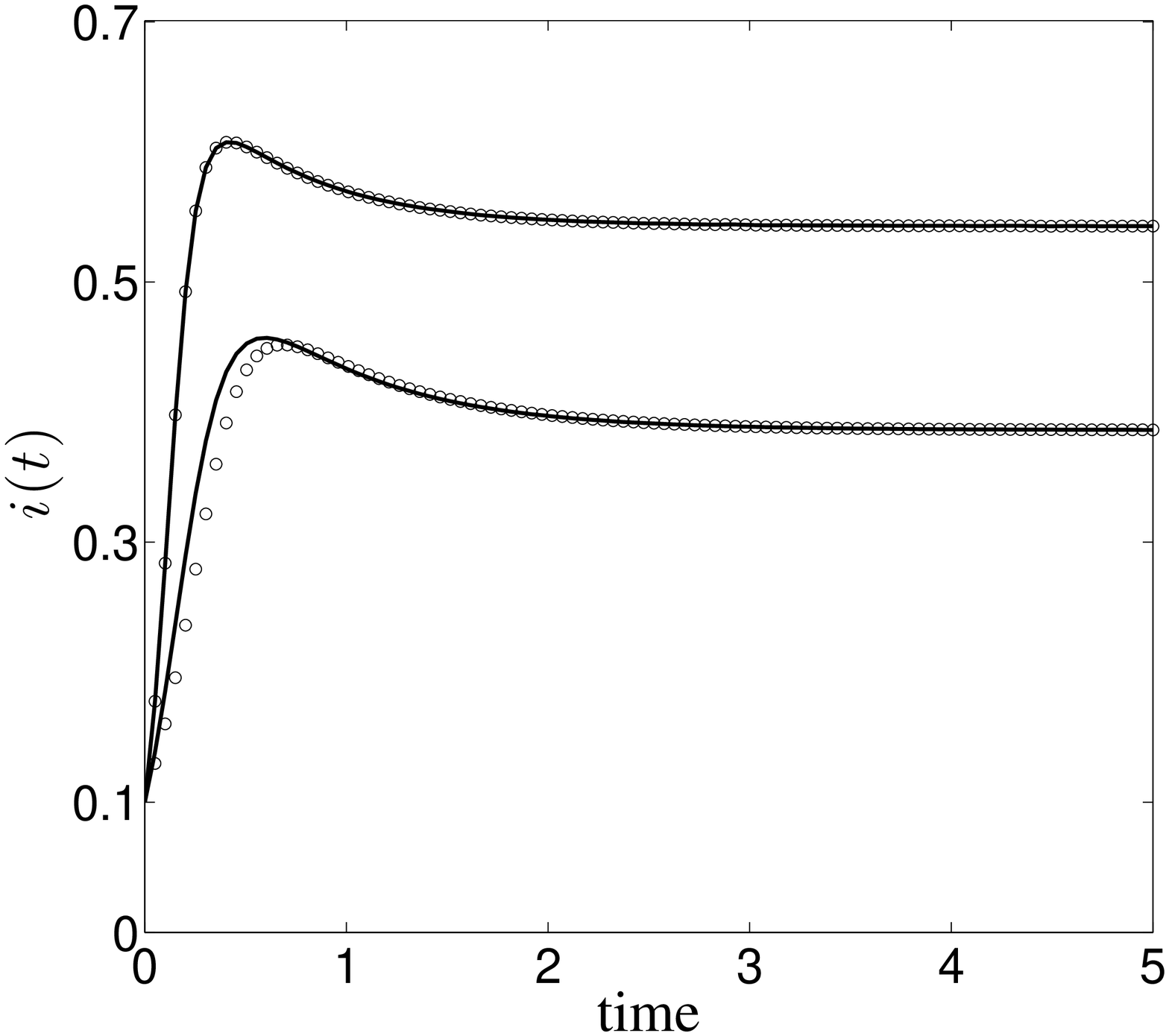}
\caption{Evolution of the fraction of infectious nodes $i(t)$ for a smaller epidemic ($\delta=4$, $\delta_a=0.5$, $\beta=12$, $\beta_a=2$ and $\kappa=4$) and 
a larger epidemic ($\delta=4$, $\delta_a=0.5$, $\beta=18$, $\beta_a=2$ and $\kappa=4$) on a regular random network with $N=1000$ nodes of degree $k=5$. 
Open circles ($\circ$) correspond to the solutions of the node-based model \eqref{markov}, with $\beta_0 = \beta/k$, $\beta_a^0  = \beta_{a}/k$, $\kappa_0=\kappa/k$. Continuous lines are the solutions of the mean-field model \eqref{edos}. For the larger epidemic the initial condition is uniform with each node having a probability $0.9$, $0.1$ and $0$ of being susceptible, infectious or aware at time $t=0$, respectively. For the smaller epidemic, the neighbours of 20 randomly chosen nodes were infected with probability $1.0$, resulting in a 10\% infectivity at time $t=0$. As proven in Lemma~\ref{equiv}, the output from the two models coincide for uniformly random initial conditions. For initially clustered infections, the mean-field model overestimates the initial epidemic growth predicted by the node-based one, although both solutions tend to the same steady state $i^*=0.38618$.}
\label{MF-vs-Intertwined}
\end{figure}

System~\eqref{edos} constitutes an extension of the SIS model on regular networks \cite{Kephart}. Remarkably, system~\eqref{edos} has been derived without summing the equations for $p_i(t)$ and $q_i(t)$ in \eqref{markov} with respect to $i$, which allow us to prove that both formulations have equivalent solutions under uniformly random initial infections, even for networks that are not fully connected ($k <N-1$). Note that, even for fully connected networks, the assumption of uniformly random initial conditions is required for a complete equivalence of the formulations, as it was observed in \cite{Mieghem} for the SIS model. In other words, we have shown that mean-field epidemic models on regular networks do not underestimate the rate of infection obtained from the full system \eqref{markov} when we restrict ourselves to uniformly random initial conditions, as it was claimed for the SIS model on regular networks with $k < N-1$ when it was compared to the corresponding full (intertwined) system (cf. Sect. VII in \cite{Mieghem}). In Figure \ref{MF-vs-Intertwined} we compare the solutions of both models under uniformly random and clustered initial conditions on networks of size $N=1000$ and degree 5. As proved in the previous lemma, the time evolution of the fraction of infectious individuals predicted by both models is exactly the same when $(p_i(0), q_i(0))=(p_0,q_0) \,\forall i$. In contrast, this figure also shows that, when we initially infect the neighbours of 20 randomly selected nodes (clustered infections), and aware individuals are not present, system \eqref{markov2} overestimates the early epidemic growth. In both models, however, the solutions tend to the same equilibrium point. 

Following these findings, we now turn our attention to the analysis of two mean-field models. Namely, the mean-field model corresponding to the information transmission model proposed in \cite{Sahneh12a, Sahneh11}, and an extended version of it, which accounts for the decay of awareness. In the following two sections, we will give a detailed bifurcation analysis of both, and we will show that the two-threshold feature (i.e. the classic invasion threshold not involving information and the die-out threshold delimiting the persistence from the eventual die-out following a minor outbreak) of the first model disappears via a degenerate transition upon including awareness decay.  Moreover, our analysis will show that the qualitative bifurcation picture from the mean-field model maps exactly to the behaviour obtained from the full model  \eqref{markov}, as detailed in \cite{Sahneh12a, Sahneh11}.


\section{Analysis of the mean-field SAIS without awareness decay} \label{SAIS-no-decay}

Let us begin with the mean-field equations obtained from \eqref{edos} with $\delta_a = 0$, namely,
\begin{equation}\label{edos-nodecay}
\left\{\begin{array}{lll}
\displaystyle \frac{ds}{dt} & = & (1-s-a)(\delta-(\kappa+\beta)s) \\
& & \\
\displaystyle \frac{da}{dt} & = & (1-s-a)(\kappa s-\beta_aa),
\end{array}\right.
\end{equation}
where $\beta = k \beta_0$, $\kappa = k \kappa_0$, and $\beta_a = k \beta^0_a$ are transmission rates per node.

Some basic features of the model are given by the following
\begin{lemma}\label{basic-nodecay}
Assume $\beta,\beta_a,\kappa, \delta > 0$. Then
\begin{enumerate}
\item[(a)] The region $R = \left\{ (s, a) \in \mathbb{R}^2 \ | \, 0 \le a+s \le 1, \, s \in [0,1] \right\}$ is positively invariant under the flow induced by system \eqref{edos-nodecay}. \\
\item[(b)] Any point on the segment $L = \left\{ (s, a) \in \mathbb{R}^2 \ | \ a+s=1, \, s \in [0,1] \right\}$ is a disease-free equilibrium. \\
\item[(c)] If  $\kappa + \beta - \beta_a \ne 0$, the trajectories passing through points $(s,a) \in R$ with $s+a < 1$ are given by
\begin{equation} \label{solution}
a(s) = C_0 \left( \delta - (\kappa + \beta) s \right)^{\beta_a/(\kappa+\beta)} + \frac{\kappa (\delta - \beta_a s)}{\beta_a(\kappa + \beta - \beta_a)}
\end{equation}
where $C_0$ is determined by the initial condition $(s(0),a(0))$ of the trajectory.
\end{enumerate}
\end{lemma}

\textit{Proof}. Statements $(a)$ and $(b)$ directly follow from the analysis of the vector field defined by \eqref{edos-nodecay}. In particular, evaluating it on the left and bottom boundaries of $R$ we obtain that
$\left. \frac{ds}{dt} \right|_{s=0} = \delta (1- a) > 0$ $\forall a \in [0,1)$ and $\left. \frac{da}{dt} \right|_{a=0} = \kappa s (1-s) > 0$ $\forall s \in (0,1)$, respectively, which means that trajectories cannot leave $R$ through these boundaries.

Statement $(c)$ follows from the fact that, for those points $(s,a)$ such that $s+a <1$, we can divide the second equation of \eqref{edos-nodecay} by the first one to obtain a non-homogeneous first order linear equation for $a(s)$, namely,
$
\frac{da}{ds} =  \frac{\kappa s - \beta_a a}{\delta-(\kappa + \beta)s}.
$
This equation can be integrated and its general solution is given by \eqref{solution}.
$\Box$

The dynamics of system \eqref{edos-nodecay} is summarized in the next theorem which, for simplicity, is proven by analysing $da/ds$ instead of using the expression of $a(s)$.

\begin{theorem} \label{GB}
Assuming $\beta_a, \beta, \kappa, \delta > 0$ with $\beta_a < \beta$, the global behaviour of the solutions of system \eqref{edos-nodecay} is given by one of the following cases:
\begin{enumerate}
\item[(a)] For $\beta \le \delta$, any trajectory inside $R$ ends up at a point of $L$. Along each trajectory, the sum $s+a$ increases monotonously towards 1.

\item[(b)] For $\beta_a < \delta < \beta$, the value $s^*_0 = \frac{\delta - \beta_a}{\beta-\beta_a}$ defines a point $(s^*_0, 1-s^*_0)$ on the segment $L$ that
splits it into two regions, $L_1$ and $L_2$, such that $$L_1 := \left\{(s,1-s)  \ | \ s^*_0 \le s \le 1 \right\}$$ defines a continuum of unstable equilibria whereas
$L_2 := L \setminus L_1$ defines a continuum of locally stable equilibria. Then,

\begin{enumerate}
\item[(i)] if $\kappa \ge \kappa^* := \frac{\beta_a(\beta-\delta)}{\delta-\beta_a}$, there are no interior equilibria and every trajectory inside $R$ tends to an equilibrium in $L_2$,

\item[(ii)] if $\kappa < \kappa^*$, there exists an equilibrium $(s^*, a^*) := \left( \frac{\delta}{\kappa + \beta}, \frac{\kappa \delta}{\beta_a(\kappa + \beta)} \right) \in R$,
which bifurcates from $(s^*_0, 1-s^*_0)$ at $\kappa = \kappa^*$. This equilibrium is asymptotically stable and attracts any trajectory inside $R$ with an initial $a < 1-s^*_0$. Those trajectories inside $R$ not tending to $(s^*, a^*)$ approach  an equilibrium in $L_2$.
\end{enumerate}

\item[(c)] For $\delta < \beta_a < \beta$, the equilibrium $(s^*, a^*)$ is globally asymptotically stable, i.e., it attracts all the trajectories inside $R$.
\end{enumerate}
\end{theorem}

\textit{Proof}. In the proof of (a) and (b), we will neglect the factor $1-s-a$ affecting both equations because, although it becomes 0 on $L$, it does not affect the slope of the vector field inside $R$ (it cancels out when we divide the equations of \eqref{edos-nodecay} to obtain $da/ds$).

To prove statement $(a)$, two different situations must be considered. First, we can have $\delta \ge k+ \beta$. In this case, trajectories inside $R$ move to the right and their slope is always greater than $-1$. More precisely,  $\frac{da}{ds} = \frac{\kappa s - \beta_a a}{\delta-(\kappa + \beta)s} > -1$ if, and only if, $\delta > \beta s + \beta_a a$ because $s \in [0,1]$. But, $\beta s + \beta_a a < (\beta - \beta_a) s + \beta_a \le \beta \le \delta$ because $a \in [0, 1-s)$ and $\beta > \beta_a$.  Therefore, the trajectories cross the straight lines $s + a = c$, $c \in (0,1)$, only once. In the second case, $\beta < \delta < \beta + k$. For $s < s^* := \delta/(\beta + k) < 1$, the same result as before follows: trajectories move to the right and $da/ds > -1$. For $s > s^*$,  trajectories move to the left and it can be shown along the same lines that the slope of the trajectories $da/ds < -1$. So, also in this second case, trajectories cross the lines $s + a = c$, $c \in (0,1)$, only once. Note that $s=s^*$ is a vertical straight line and corresponds to the nullcline $\frac{ds}{dt}=0$ and, hence, trajectories cannot cross it. Therefore, in both cases the sum $s + a$ tends to 1 monotonously along trajectories because of the invariance of $R$ (Lemma \ref{basic-nodecay}). That is, trajectories tend monotonously to $L$, which means that no minor outbreaks are possible.

\begin{figure}[ht]
\begin{tabular}{cc}
\hspace{-1cm}
	\includegraphics[scale=0.35]{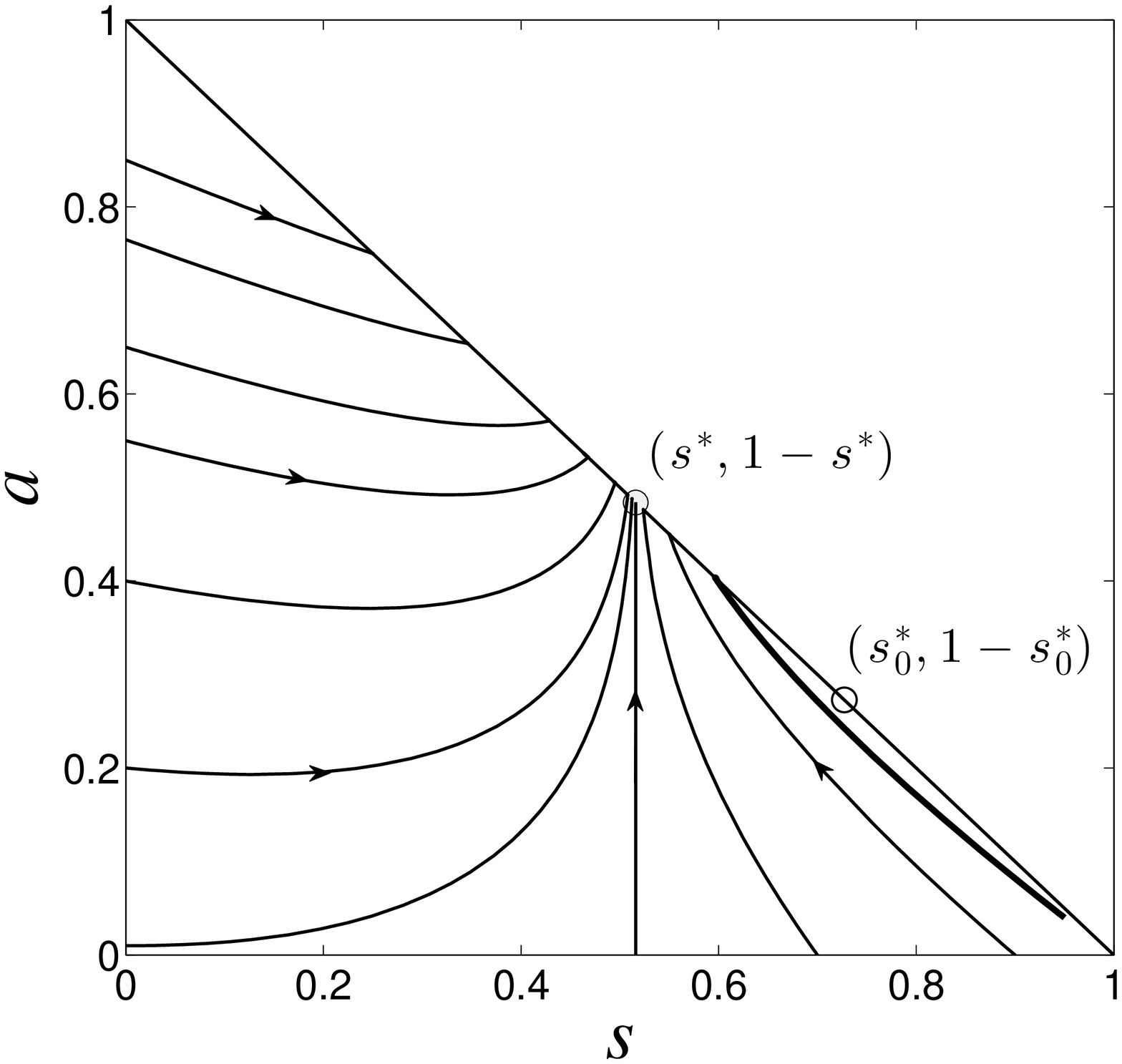}
&
	\includegraphics[scale=0.35]{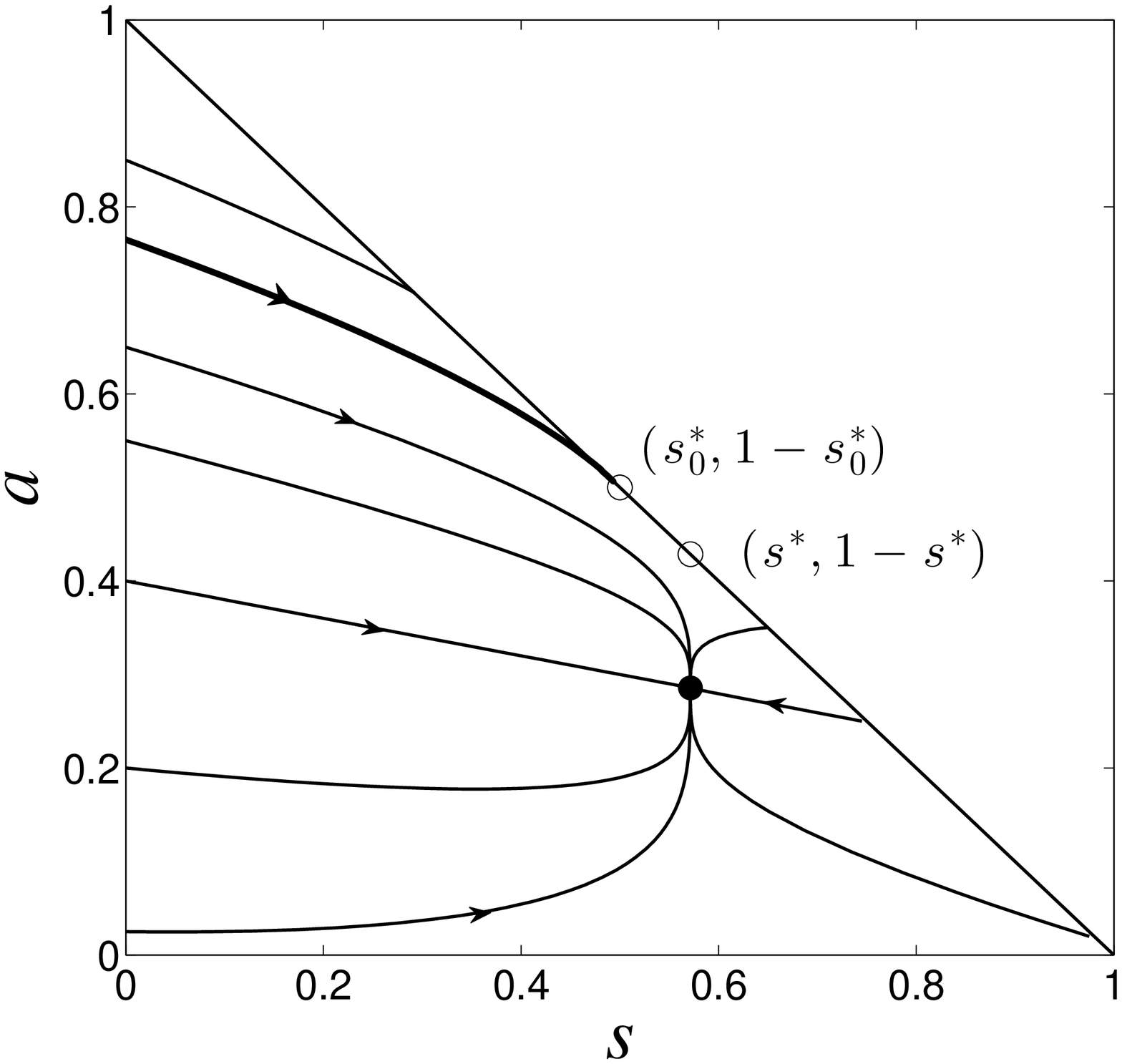}
\end{tabular}
\caption[Phase]{Phase portrait of system \eqref{edos-nodecay} for $\beta_a < \delta < \beta$. Left panel: $\kappa \ge \kappa^*$. The thick line corresponds to a minor outbreak. The end point of the attracting boundary $L_2$ is $(s^*_0,1-s^*_0) = (0.7273, 0.2727)$. Right panel: $\kappa < \kappa^*$. The thick line is the trajectory ending up at $(s^*_0, 1-s^*_0)=(0.5, 0.5)$ that limits from above the basin of attraction of the endemic equilibrium $(s^*,a^*)=(0.5714, 0.2857)$, here represented by a solid dot. Parameters: $\delta=4$, $\beta_a=2$, and $\beta=4.75$, $k=3$ (left) and $\beta=6$, $\kappa=1$ (right).}
\label{fig:Case2}
\end{figure}

In statement $(b)$, the existence of the value $s^*_0$ for $s$ follows from the analysis of the slope of the trajectories when they approach $L$, and from the direction of the vector field. In particular, we are interested in those points on $L$ where the vector field is parallel to $L$, that is, where
$ \left. \frac{da}{ds} \right|_{a=1-s} = -1$. From this expression, we obtain that $s^*_0:= \frac{\delta - \beta_a}{\beta-\beta_a}$ is the only value of $s$ for which $\frac{da}{ds} = -1$ on $L$. This point separates the region $L_2$, where trajectories approach $L$, from the region $L_1$ where trajectories depart from the immediate vicinity of $L$.
For $\kappa \ge \kappa^*$, trajectories move to the right and $\left. \frac{da}{ds} \right|_{L} > -1$  for all  $s \in [0, s^*)$, whereas trajectories move to the left and $\left. \frac{da}{ds} \right|_L < -1$  for all $s \in (s^*, s^*_0)$, with $s^*$ being the location of the vertical nullcline. Therefore, $L_2 = \{(s, 1-s) \, | \, s \in [0, s^*_0)\}$ and, hence, $L_1 = L \setminus L_2 = \{(s, 1-s) \, | \, s \in (s^*_0,1)\}$, with trajectories moving to the left and $\left. \frac{da}{ds} \right|_{L_1} > -1$.
At $(s, a) = (s^*, 1-s^*)$, the nullcline $\frac{ds}{dt}=0$ and $L$ intersect each other and, hence, the vector field is vertical and upwards on this nullcline. So, those trajectories departing from the vicinity of $L_1$ end up at $L_2$ within the region defined by $s^* < s < s^*_0$ (see left panel in Figure \ref{fig:Case2}). These trajectories cross the lines $s + a = c$ with $c \approx 1$ twice.

For $\kappa < \kappa^*$,  there exists a unique interior equilibrium $(s^*, a^*)$, with $a^*=a(s^*)$. Moreover, trajectories move to the right and $\left. \frac{da}{ds} \right|_L > -1$ for all $s \in [0, s^*_0)$, i.e., $L_2 = \{(s, 1-s) \, | \, s \in [0, s^*_0)\}$. Trajectories leaving the vicinity of $L_1 = \{(s, 1-s) \, | \, s \in (s^*_0,1)\}$ move to the right with $\left. \frac{da}{ds} \right|_L < -1$ for all $s \in (s^*_0, s^*)$, and move to the left with $\left. \frac{da}{ds} \right|_{L} > -1$ for all $s \in (s^*, 1)$. From Poincar\'e-Bendixson theorem, it follows that all of them approach the endemic equilibrium because periodic orbits are not possible since the vertical nullcline $\frac{ds}{dt}=0$ is a vertical straight line (see right panel in Figure \ref{fig:Case2}).

Statement $(c)$ also follows from Poincar\'e-Bendixson theorem because, as before, there exists a unique interior equilibrium $(s^*, a^*)$, no periodic orbits are possible within $R$, and now trajectories depart from the immediate vicinity of the whole segment $L$ (because $s^*_0 < 0$).
$\Box$

Note that, in case $(b)$ with $\kappa \ge \kappa^*$, those trajectories departing from the vicinity of $L_1$ correspond to the ocurrence of minor outbreaks that eventually disappear as these trajectories approach $L_2$.  In Table~\ref{SIASNoDec} we present a summary of the possible dynamical behaviours of the model.

\vspace{0.5cm}

\begin{table}[h]
\centering
\begin{tabular}{|l|l|l|l|l|}
\hline
\multirow{2}{*}{$\beta_a < \beta < \delta$} & \multicolumn{2}{|l|}{$\beta_a < \delta < \beta$}  &   \multirow{2}{*}{$\delta < \beta_a < \beta$}  \\  \cline{2-3}
                                                                 & $\kappa  > \kappa^{*}$   & $\kappa < \kappa^{*}$       &                \\ \hline
\multirow{2}{*}{}
                \vtop{\hbox{\strut Trajectories inside $R$}\hbox{\strut tend to different}\hbox{\strut disease-free states}\hbox{\strut $(s_{\infty}, 1-s_{\infty})$.}}                    &
                \vtop{\hbox{\strut Minor outbreaks}\hbox{\strut followed by die}\hbox{\strut out. No endemic}\hbox{\strut equilibrium.}}        &
                \vtop{\hbox{\strut The endemic}\hbox{\strut equilibrium}\hbox{\strut is locally}\hbox{\strut stable.}}       &
                \vtop{\hbox{\strut The endemic}\hbox{\strut equilibrium}\hbox{\strut is globally}\hbox{\strut stable.}}      \\ \hline
\end{tabular}
\vspace{0.3cm}
\caption{Summary of model behaviours, with the richest dynamical features in the $\beta_a < \delta < \beta$ regime where there exists a continuum of stable disease-free equilibria. The die-out threshold is defined by $\kappa=\kappa^{*}:=\beta_{a} (\beta-\delta) / (\delta-\beta_a)$. The points of the form $\left\{ (s, a) \in \mathbb{R}^2 \ | \ a+s=1, \, s \in [0,1] \right\}$ are always equilibria.}
\label{SIASNoDec}
\end{table}

\begin{figure}[ht]
\centering
\begin{tabular}{cc}
	\includegraphics[scale=0.30]{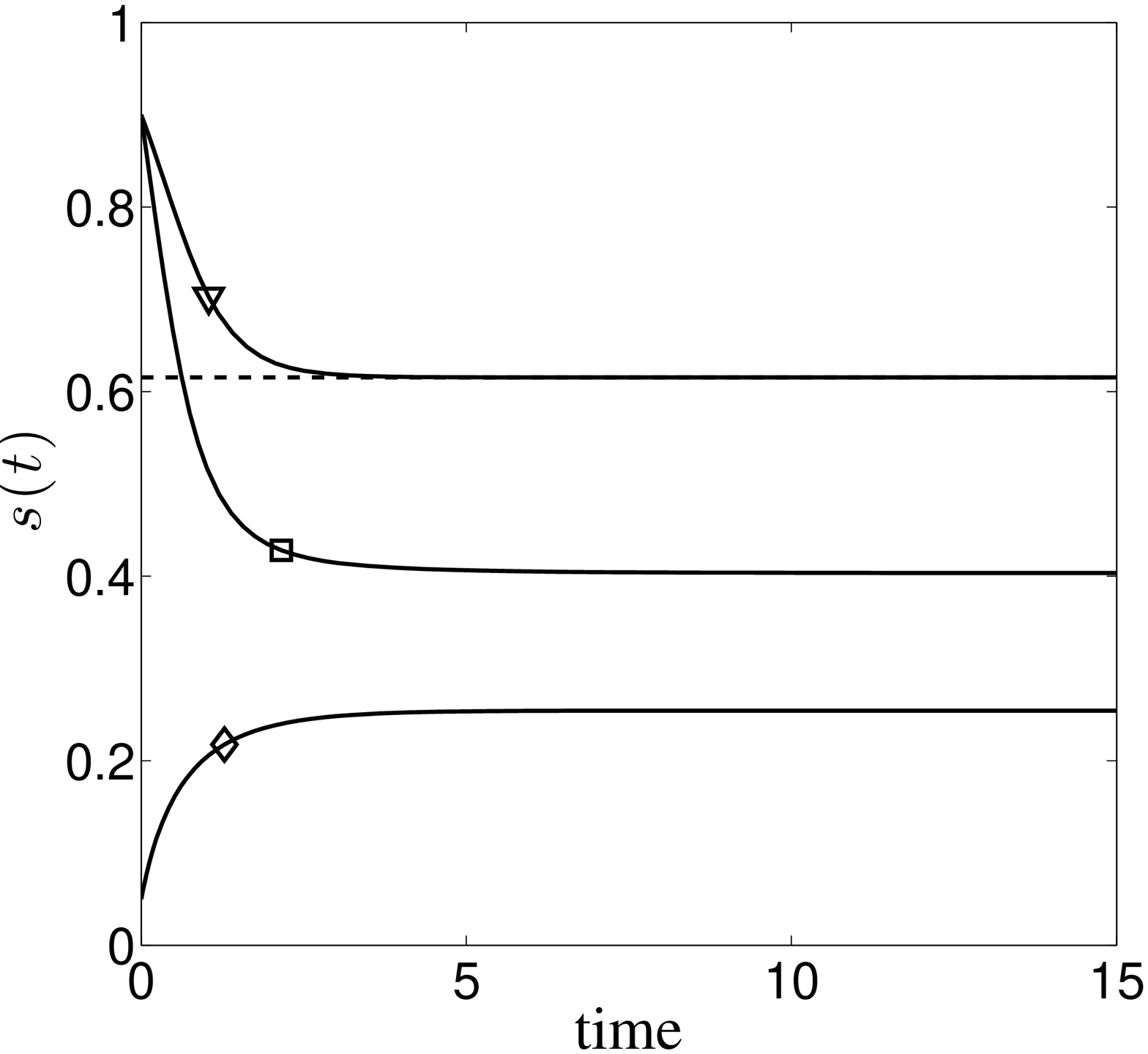}
	&
	\includegraphics[scale=0.30]{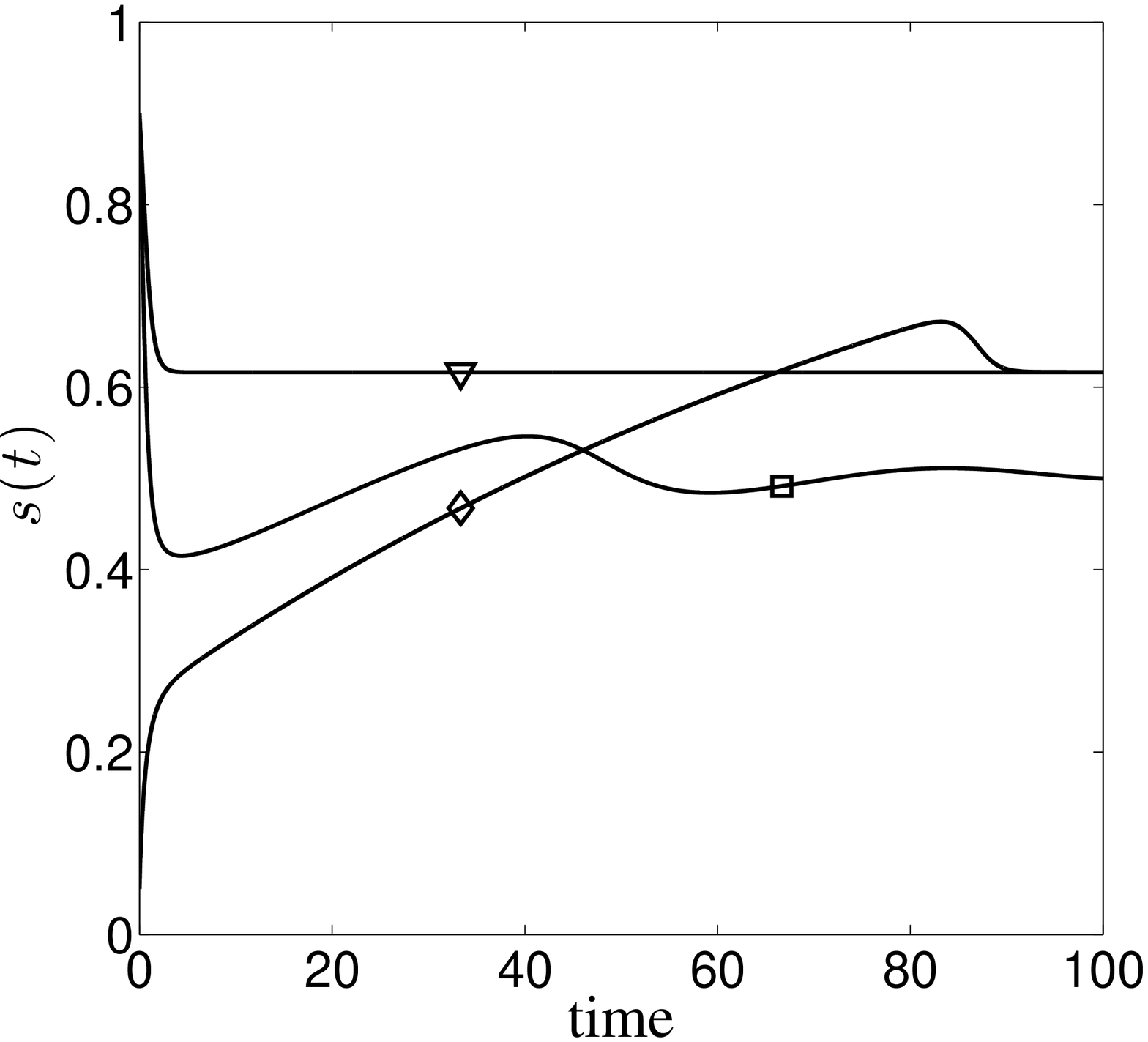}
\\	
	\includegraphics[scale=0.30]{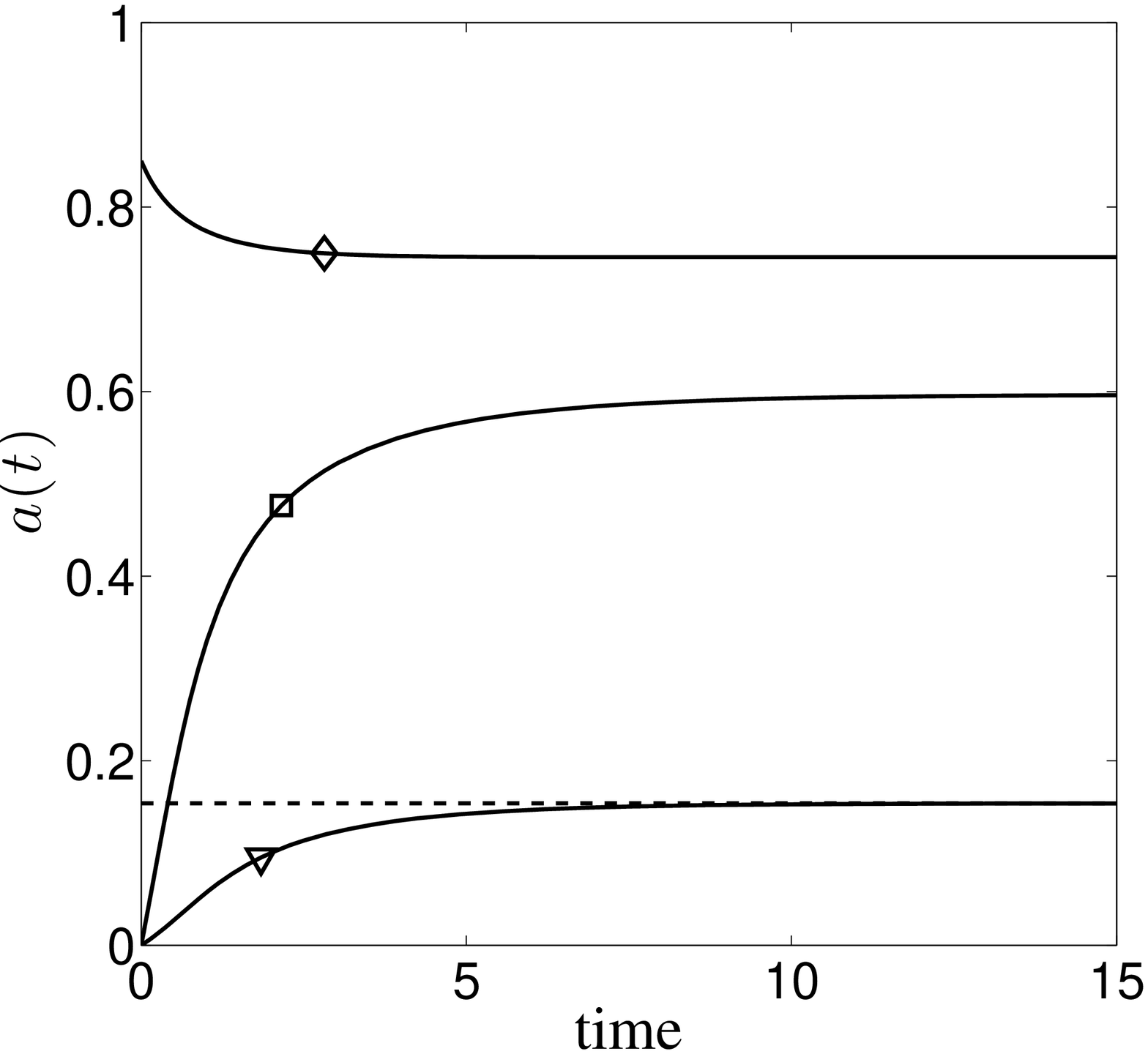}
	&
	\includegraphics[scale=0.30]{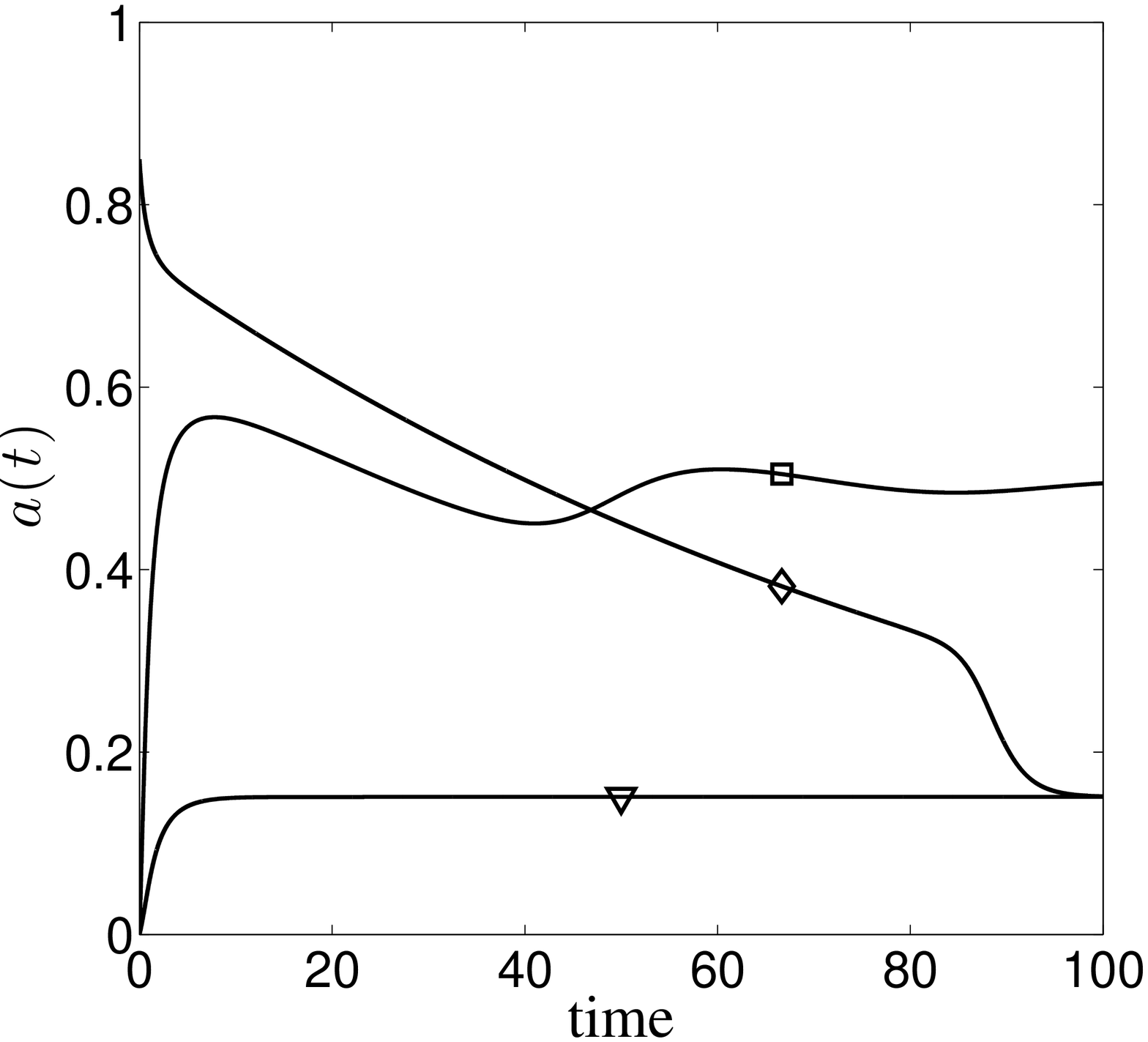}
\\
	\includegraphics[scale=0.30]{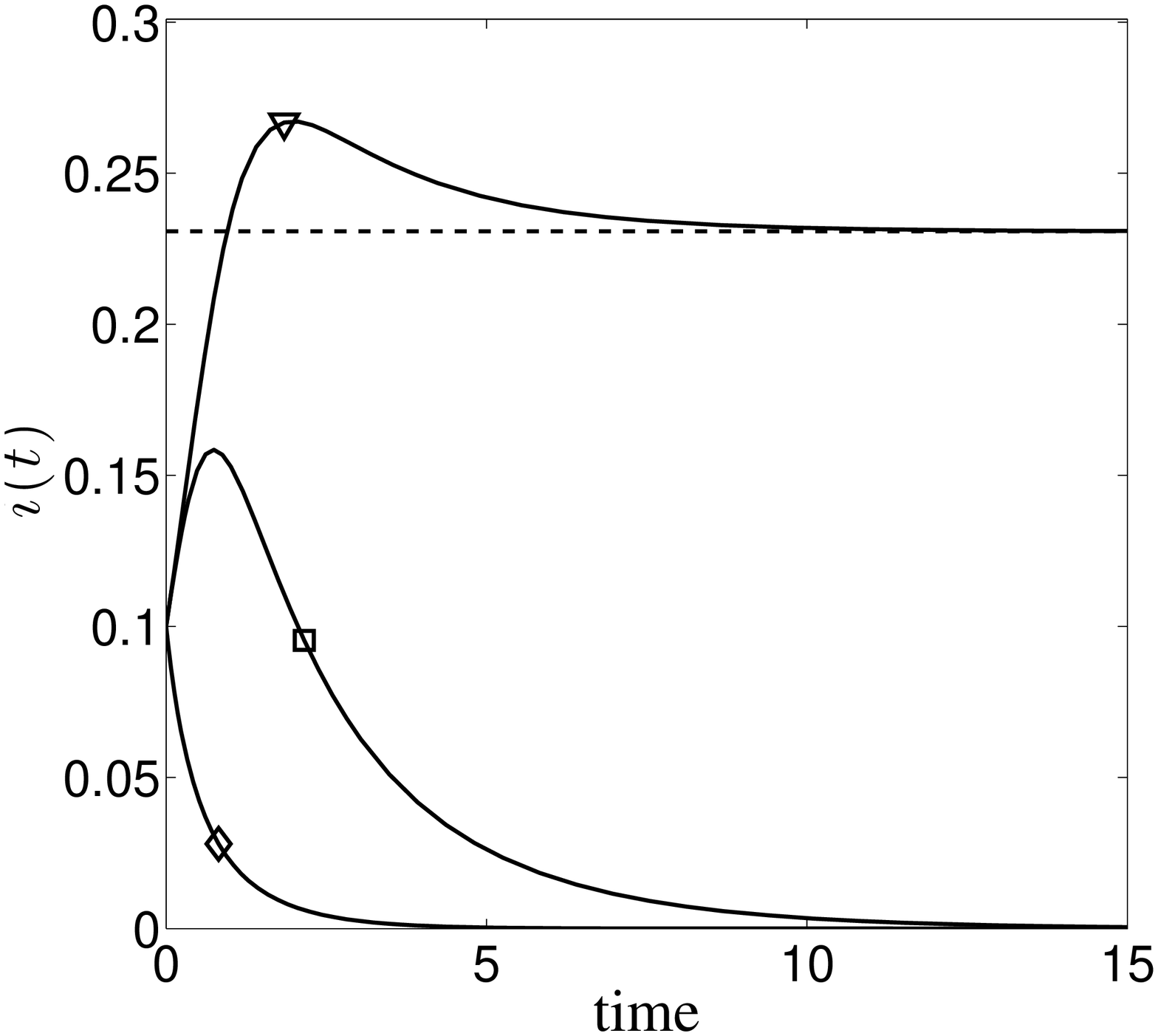}
	&
	\includegraphics[scale=0.30]{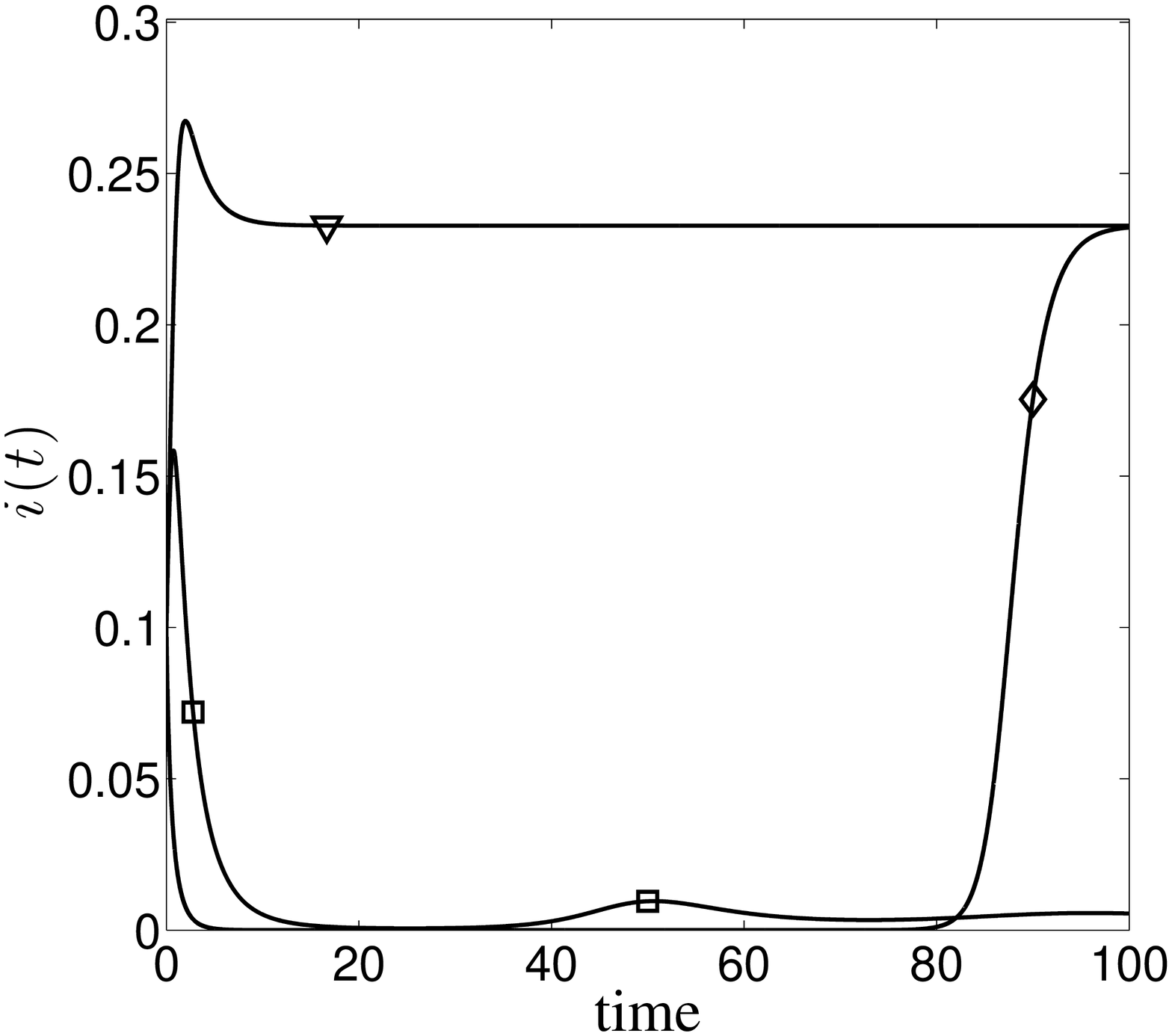}
\end{tabular}
\caption{Examples of typical time evolution of solutions to systems \eqref{edos-nodecay} (left) and \eqref{edos} (right) with a very small rate of awareness decay ($\delta_a = 0.01$)  for the cases in the two middle columns of Table~\ref{SIASNoDec}, i.e., below and above the second threshold for $\delta_a=0$. The following cases are covered: 
($\Box$) $\delta=4$, $\beta=6$, $\beta_a=2$, $\kappa=4$ with $\kappa^{*}=2<\kappa$;
($\bigtriangledown$) $\delta=4$, $\beta=6$, $\beta_a=2$, $\kappa=0.5$ with $\kappa^{*}=2>\kappa$;
and ($\Large \Diamond$) same as ($\bigtriangledown$) but with $s(0)=0.05$, $a(0)=0.85$, $i(0)=0.1$. 
All other initial conditions are set at: $s(0)=0.90$, $a(0)=0$, and $i(0)=0.1$. The dashed line in the left panels corresponds to the endemic equilibrium $(s^*,a^*)$ as given in Theorem \ref{GB}. Trajectories $\bigtriangledown$ and $\Large \Diamond$ do not tend to the same limit in the left panels because $(s^*,a^*)$ is not globally stable. Notice the longer transient behaviour of solutions in the right panels which is due to the very small value of $\delta_a$.  
}
\label{fig:evolution_of_cases}
\end{figure}

From this analysis of the dynamics of system \eqref{edos-nodecay}, the nature of the die-out epidemic threshold $\kappa = \kappa^*$ found by Scoglio et al. in \cite{Sahneh12a, Sahneh11} becomes clearer. Its occurrence is based on the existence of a continuum of equilibria, part of which (the one nearest to a purely susceptible population) turns out to be unstable. This happens when the recovery rate $\delta$ becomes lower than the infection rate $\beta$ of a susceptible individual, but it is still higher than the infection rate $\beta_a$ of an aware individual. In such circumstances, there can be minor outbreaks if most of the population is susceptible but, if the awareness rate is high enough ($\kappa > \kappa^*$), the creation of aware individuals does not allow for infectious individuals to be present at equilibrium. Eventually, the population will be composed only of susceptible and aware individuals, defining a new kind of disease-free state. In Figure \ref{fig:evolution_of_cases} (left column) we compare the time evolution of susceptible, infectious and aware individuals for different parameters values to illustrate the nature of these minor outbreaks. 

We also show in Figure~\ref{fig:evolution_of_cases} (right column) solutions of system~\eqref{edos} for a very small rate of awareness decay ($\delta_a=0.01$) and the same values for the rest of parameters and the same initial conditions as in the left panels. We can see that, for very small values of $\delta_a$, minor outbreaks occurring in system \eqref{edos-nodecay} when $\beta_a < \delta < \beta$ and $\kappa > \kappa^*$ (see also left panel of Figure~\ref{fig:Case2}) are replaced by damped oscillations towards an interior equilibrium with a very low prevalence of the disease (compare curves marked by squares ($\Box$) in both columns). One can also see a trajectory with a high initial fraction of aware individuals tending to a disease-free equilibrium on the boundary $s+a=1$ if $\delta_a=0$, whereas, for the same intial condition and $\delta_a > 0$, the trajectory tends to an endemic equilibrium with a significant prevalence of the disease after a long period of time with a extremely low prevalence (compare solutions marked with diamonds ($\Diamond$) in Figure~\ref{fig:evolution_of_cases}). Such different qualitative behaviours of the solutions is a consequence of the structural instability of the (continuously differentiable) flow induced by system \eqref{edos-nodecay}. In the last case ($\Diamond$), system \eqref{edos-nodecay} has an endemic equilibrium which is not globally stable because of the existence of stable equilibria on the boundary, whereas system \eqref{edos} has a globally stable endemic equilibrium. Note that system~\eqref{edos-nodecay} is a two-dimensional system defined on the region $R$, which is a compact subset of $\mathbb{R}^2$, and has a continuum of equilibria, the boundary $L$. Therefore, the structural instability of the flow associated to system~\eqref{edos-nodecay} is guaranteed by Peixoto's theorem \cite{Perko}.   

\begin{figure}[ht]
\begin{tabular}{ll}
\hspace{-1cm}
	\includegraphics[scale=0.4]{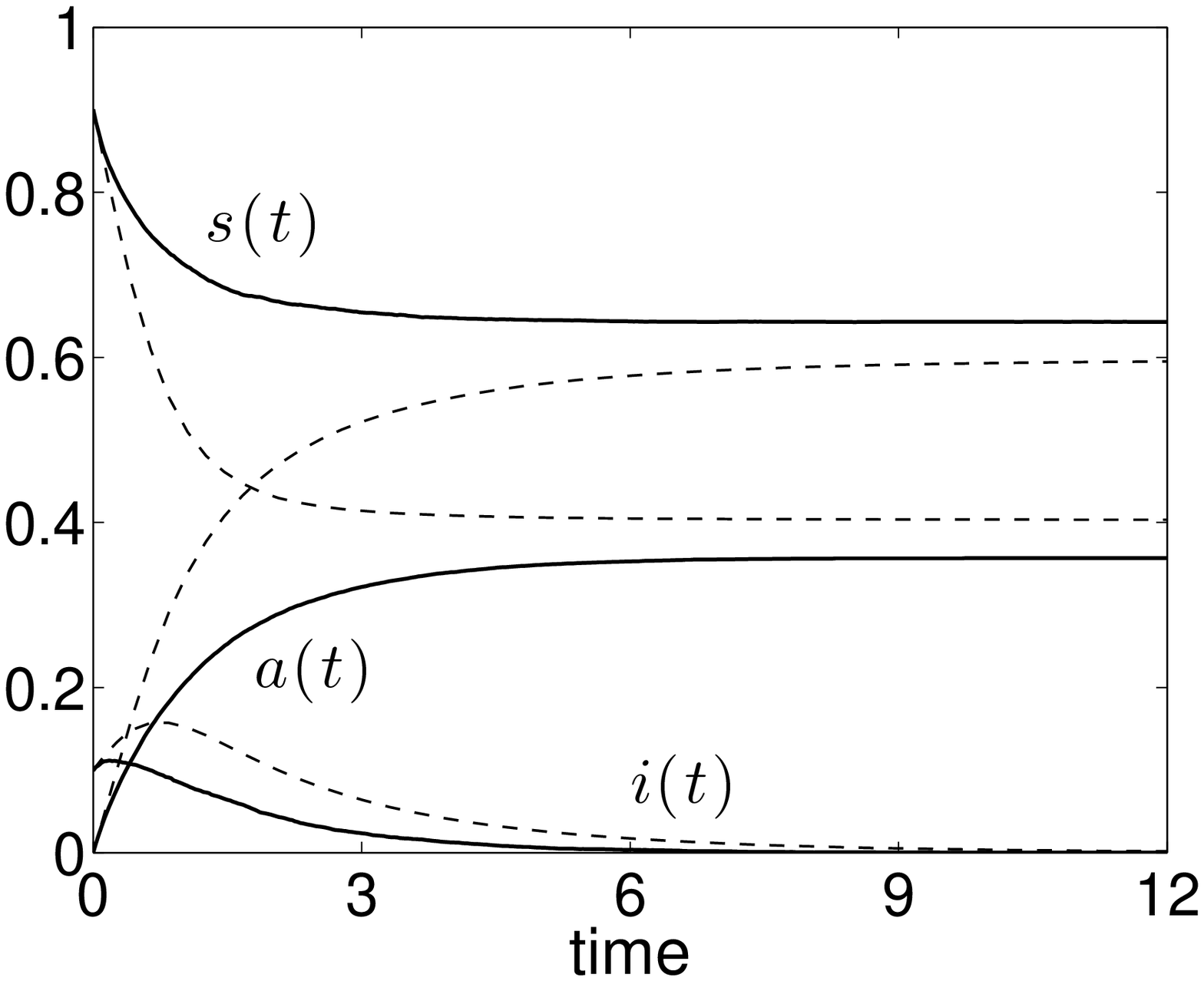}
&
	\includegraphics[scale=0.4]{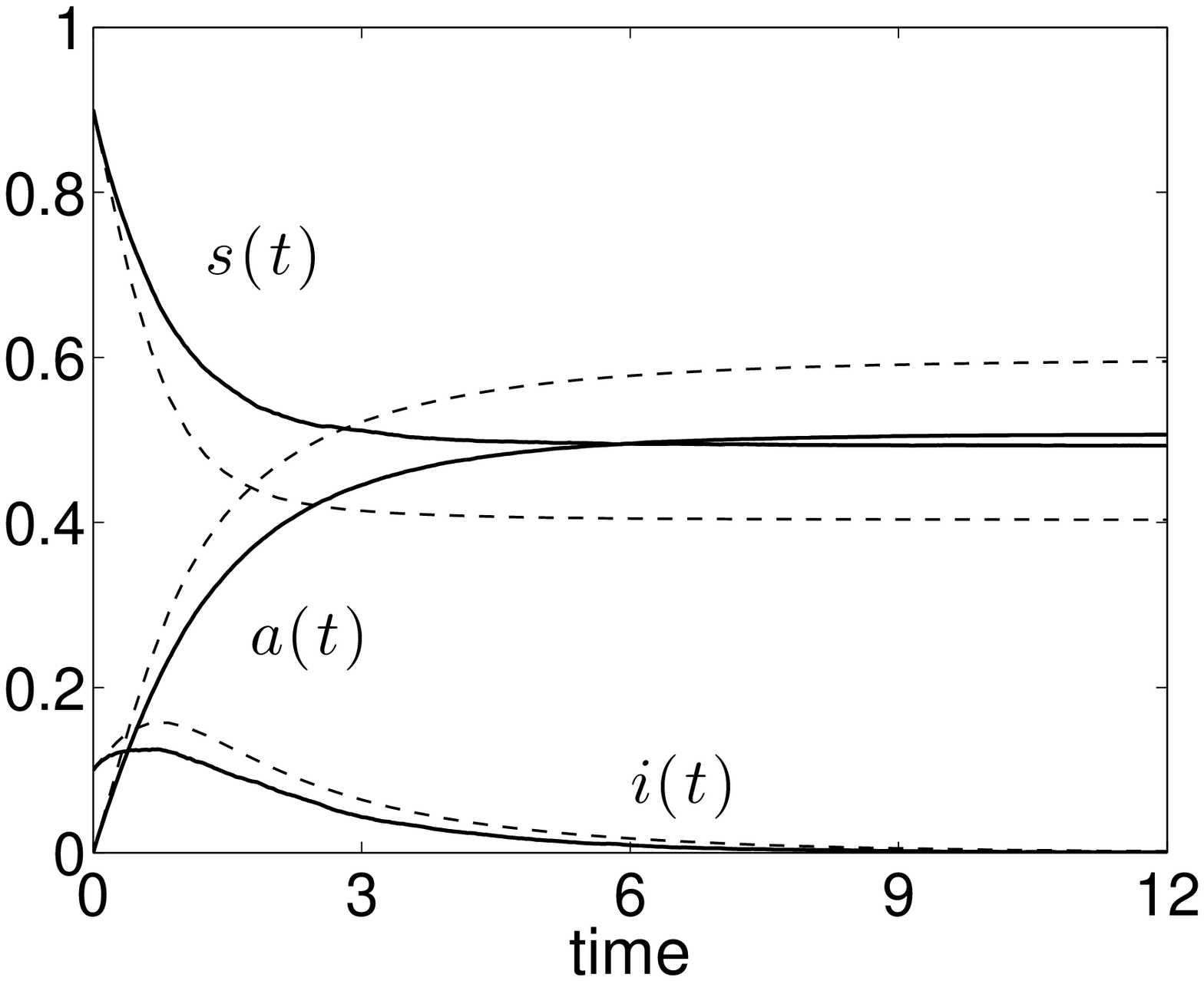}
\\
\hspace{-1cm}
	\includegraphics[scale=0.4]{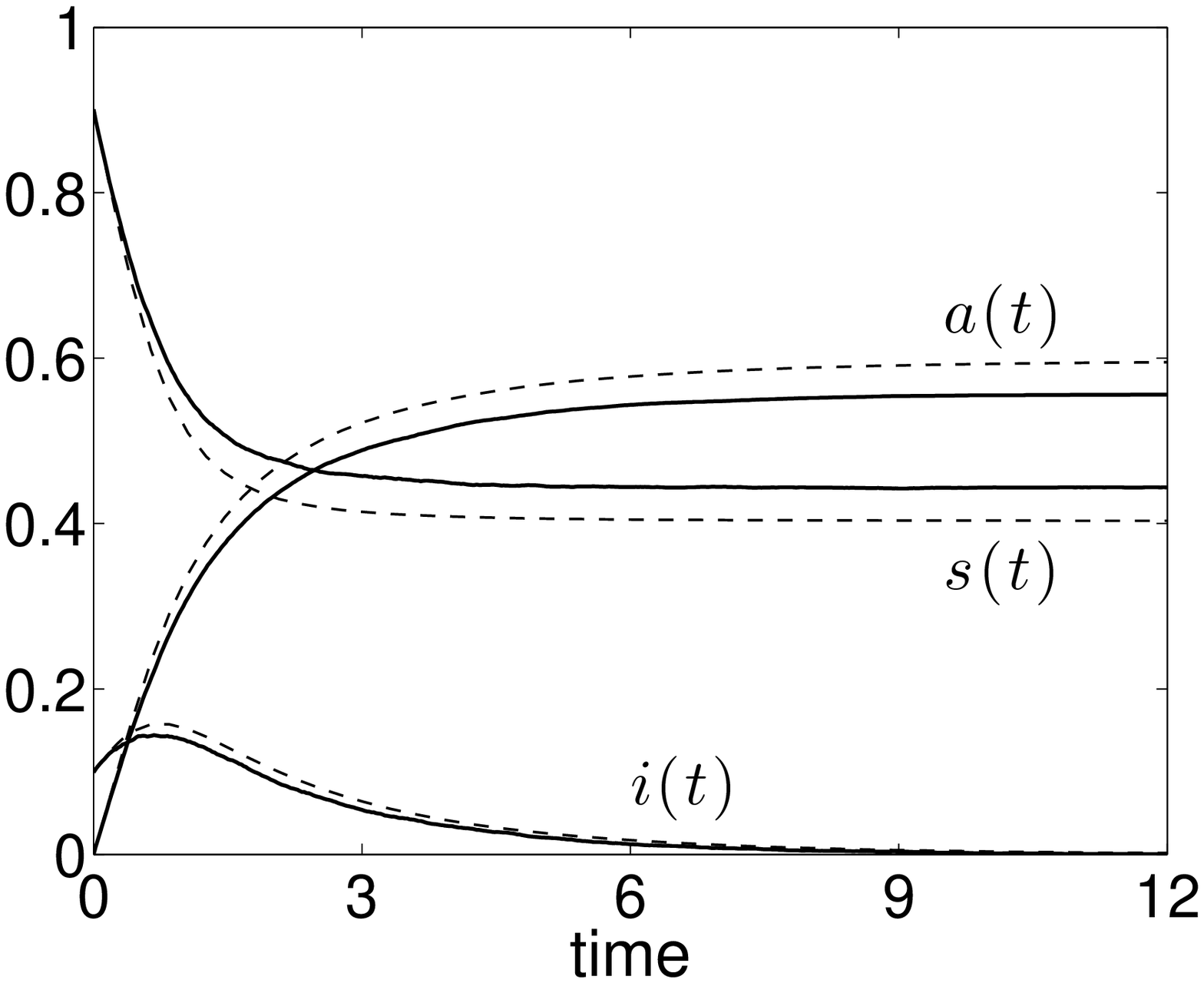}
&
	\includegraphics[scale=0.4]{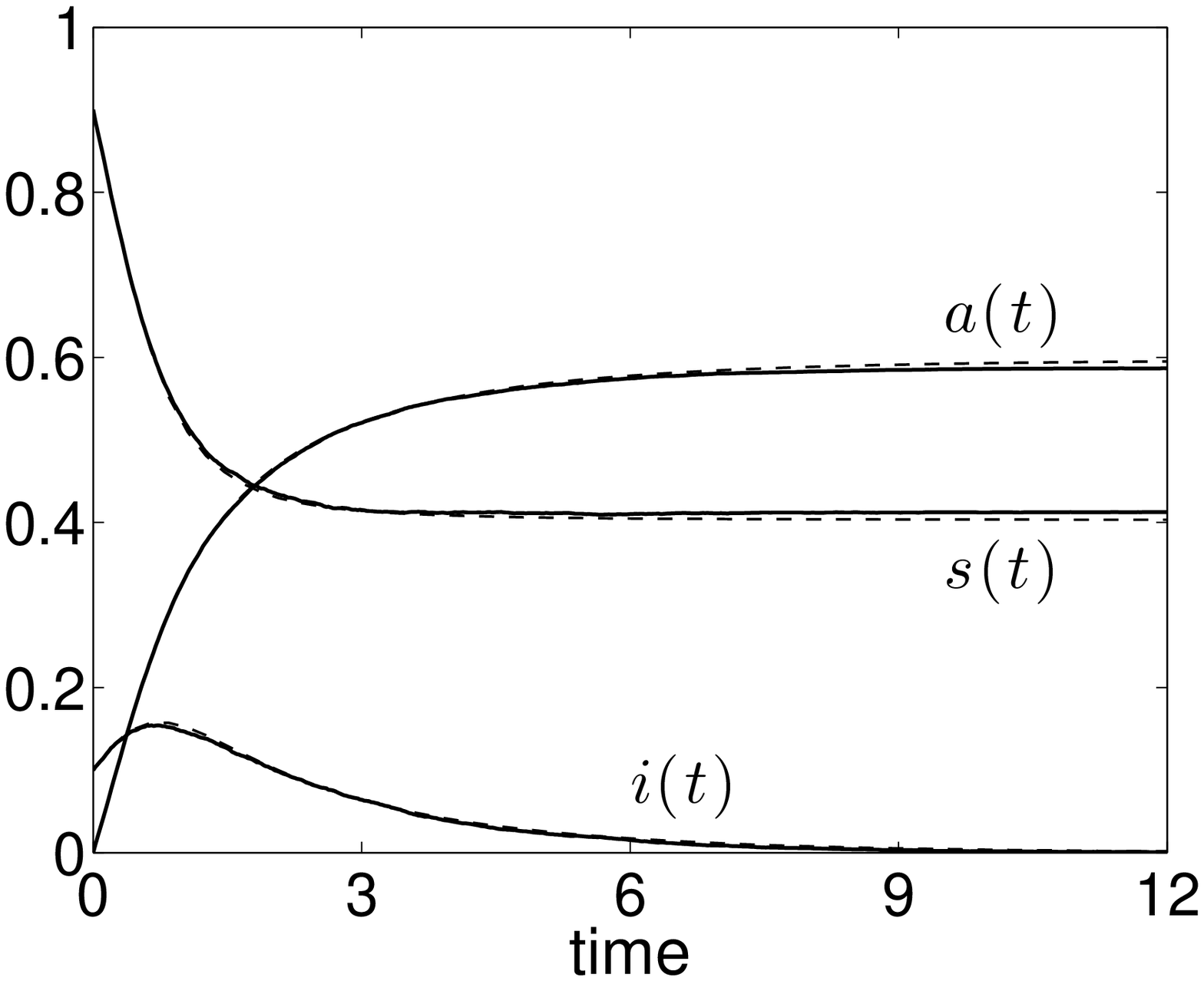}
\end{tabular}
\caption{Evolution of the fraction of infectious ($i$), susceptible ($s$), and aware ($a$) individuals with no awareness decay when a minor outbreak occurs ($\beta_a < \delta < \beta$ and  $\kappa^* < \kappa$). Dashed lines: solutions to system~\eqref{edos-nodecay} with initial condition $s(0)=0.90$, $i(0)=0.1$ and $a(0)=0$.
Solid lines: stochastic simulations over regular random networks of size 1000 and degree 5 (top left), 10 (top right), 20 (bottom left), and fully connected (bottom right). Simulation outputs averaged over 100 runs with 10\% of randomly infected individuals and 90\% of susceptible ones at $t=0$. Parameters: $\delta=4$, $\beta=6$, $\beta_a=2$, $\kappa=4$}.
\label{fig:comparison}
\end{figure}

Finally, in Figure~\ref{fig:comparison} we check the accuracy of the model when a minor outbreak is predicted by comparing its solutions with the output of continuous-time stochastic simulations of an epidemic on regular random networks of size 1000. In each panel, the time evolution of the proportion of susceptible, aware, and infectious corresponds to the average over 100 realizations. In each realization, the initially infectious cases are determined by choosing 100 individuals uniformly at random. Such a uniform initial condition is required for the equivalence among model formulations \eqref{markov} and \eqref{markov2}. As expected, the figure shows an overestimation of the fraction of aware and infectious individuals, which decreases with the network degree and vanishes in fully connected networks. The homogeneous mixing assumed in mean-field epidemic models is violated in networks, especially in those with a low nodal degree, due to the development of spatial correlations between disease status of individuals which reduce infection transmission.  


\section{Analysis of the mean-field SAIS with awareness decay} \label{SAIS-with-decay}

Now we will analyze the behaviour of the solutions of \eqref{edos} in $R$. In contrast to the previous model, note that $(s,a)=(1,0)$ is an isolated equilibrium of \eqref{edos} and corresponds to the unique disease-free equilibrium (DFE) of this system. As before, let
$R = \left\{ (s, a) \in \mathbb{R}^2 \ | \, 0 \le a+s \le 1, \, s \in [0,1] \right\}$ and $L = \left\{ (s, a) \in \mathbb{R}^2 \, | \, a+s=1, \, s \in [0,1] \right\}$. The next result establishes basic facts about this system.

\begin{lemma}\label{basic}
Assume $\beta_a, \beta, \kappa, \delta, \delta_a > 0$. Then,
\begin{enumerate}
\item[(a)] The segment $L$ is positively invariant under the flow induced by system \eqref{edos} and the trajectory of every point $(a,s) \in L$ tends to the DFE as $t\rightarrow \infty$.
\item[(b)] The region $R$ is positively invariant under the same flow. \\
\end{enumerate}
\end{lemma}

\textit{Proof}.
To prove the invariance of $L$, note that $\left. \frac{ds}{dt} \right|_L = - \left. \frac{da}{dt} \right|_L = \delta_a a$. This means that $\left. \frac{da}{ds} \right|_L = -1$, that is, any trajectory of \eqref{edos} with an initial condition on $L$ tends to $(1,0)$ along $L$.

On the other hand, we have that  $\left. \frac{ds}{dt} \right|_{s=0} > 0$, while $\left. \frac{da}{dt} \right|_{a=0} >0$ for $s \ne 0, 1$. In consequence, the vector field induced by \eqref{edos} on the lines $\{(0,a)\in R\}$ and $\{(s,0)\in R\}$ points inside $R$. Together with Lemma~\ref{basic} and the invariance of $L$, this implies that $R$ is a positively invariant region.
$\Box$

The first claim of Lemma~\ref{basic} has an obvious interpretation: if $a+s=1$, then $i=0$. In such a case, it is clear from the evolution rules \eqref{rules}
that no new aware/infectious individuals will arise, while every aware individual will become susceptible after a large enough time ($\delta_a > 0$).

Now we give a simple lemma that will help us to guarantee the uniqueness of an endemic equilibrium.

\begin{lemma}\label{rectapc}
Every equilibrium point of the system \eqref{edos} different from DFE belongs to the straight line given by
\[ a_0(s):=\frac{\delta-\beta s}{\beta_a}. \]
\end{lemma}
\textit{Proof}.
Setting $ds/dt=da/dt=0$ yields
\[ \begin{array}{lll}
\delta_a a & = & (1-s-a)((\kappa+\beta)s-\delta) \\ & & \\
\delta_a a & = & (1-s-a)(\kappa s-\beta_a a).
\end{array} \]
By Lemma~\ref{basic}, an equilibrium point $(s,a)\ne(1,0)$ does not belong to the straight line $L$. Hence, $1-s-a\ne0$ and the previous equations
amount to $(\kappa+\beta)s-\delta=\kappa s-\beta_a a$, which defines $a$ as a function of $s$, that is, $a_0(s)$.
$\Box$

The asymptotic behaviour of the trajectories of system \eqref{edos} is specially simple in the case $\delta\ge\beta$:

\begin{theorem}\label{deltagebeta}
Assume $\beta_a, \beta, \kappa, \delta > 0$ with $\beta_a < \beta$ and $\delta \ge \beta$. Then, the region $R$ contains no equilibrium points of \eqref{edos} different from DFE which is globally asymptotically stable.
\end{theorem}

\textit{Proof}.
Since $\beta_a<\beta$ by hypothesis, the slope $-\beta/\beta_a$ of the line $a_0(s)$ is less than $-1$. Note also that
$a_0(\delta/\beta)=0$. In consequence, if $\delta/\beta\ge1$ the line $a_0(s)$ does not intersect the region $R$. Then, by Lemma~\ref{rectapc},
there cannot be other equilibrium points in $R$ different from DFE.

Linearizing system \eqref{edos} about an arbitrary point $(s,a)$ it follows that the Jacobian matrix of system \eqref{edos} is
\[ J(s,a)=\left(\begin{array}{cc}
-(\delta-(\kappa+\beta)s)-(\kappa+\beta)(1-s-a) & -\delta+\delta_a+(\kappa+\beta)s \\
\\
-(\kappa s-\beta_aa)+\kappa(1-s-a) & -(\kappa s-\beta_aa)-\beta_a(1-s-a)-\delta_a
\end{array}\right). \]
At the DFE, $(s,a)=(1,0)$ and we get
\[ J(1,0) = \left(
\begin{array}{cc}
\kappa+\beta-\delta & \kappa+\beta-\delta+\delta_a
\\
-\kappa & -\kappa-\delta_a
\end{array}
\right). \]

The eigenvalues of $J(1,0)$ are $\lambda_1=-\delta_a<0$ and $\lambda_2=\beta-\delta$, with associated eigenvectors $v_1=(1,-1)$ and $v_2=(\delta-\delta_a-(\kappa+\beta),\kappa)$, respectively. So,  DFE is asymptotically stable (a stable node) for $\delta>\beta$.

Finally, the non-existence of endemic equilibria implies that no periodic orbit lies in $R$ and, hence, the global stability of the DFE follows by the invariance of $R$ and Poincar\'e-Bendixson theorem.
$\Box$

Theorem~\ref{deltagebeta} fully specifies the behavior of system \eqref{edos} when $\delta\ge\beta$. Therefore, from now on we will focus on the (more interesting) case $\delta<\beta$. Before that, we give some properties of the nullclines that will help us to establish the uniqueness of an endemic equilibrium and the global behaviour of solutions.

The vertical nullcline $ds/dt=0$ of system \eqref{edos} is given by the curve
\begin{equation}
\label{a1}
a_1(s):=\frac{(1-s)(\delta-(\kappa+\beta)s)}{\delta-\delta_a-(\kappa+\beta)s},
\end{equation}
well defined for $s\ne(\delta-\delta_a)/(\kappa+\beta)$. The next result summarizes some properties of the curve $a_1(s)$.

\begin{lemma}\label{propa1}
Assume $ \beta, \kappa, \delta, \delta_a > 0$. Then
\begin{enumerate}
\item $a_1(s)=0$ only when $s=1$ or $\displaystyle s=\frac{\delta}{\kappa+\beta}$.
\item If $\delta > \delta_a$, then $a_1(s) > 1-s$ $\displaystyle \forall s \in \left[0, \frac{\delta-\delta_a}{\kappa+\beta} \right)$. So, the graphic of $a_1(s)$ lies above $R$ for $s \in [0,1]$ if $\delta - \delta_a > \kappa + \beta$.
\item $a_1(s)<0$  $\displaystyle \forall s \in \left( \max\left\{0,\frac{\delta-\delta_a}{\kappa+\beta}\right\},\frac{\delta}{\kappa+\beta} \right)$.
\item  $0 \le a_1(s) \le 1-s$ $\displaystyle \forall  s \in \left[\frac{\delta}{\kappa+\beta}, 1 \right]$. That is, the graphic of $a_1(s)$ lies partially inside $R$
for $\delta < \kappa + \beta$.
\item If $\delta < \kappa + \beta$, then $\displaystyle a_1' \left( \frac{\delta}{\kappa+\beta} \right) > 0$ and $-1 < a_1'(1) < 0$
\item If $\delta < \kappa + \beta$, then $a_1(s)$ has a unique extremum in $\displaystyle \left( \frac{\delta}{\kappa+\beta},1 \right)$, which is  the global maximum of $a_1(s)$ inside $R$.
\end{enumerate}
\end{lemma}

\textit{Proof}.
All the statements (1)--(5) follow from straightforward computations. Let us prove (6). For $\delta < \kappa + \beta$, the equation $a_1'(s)=0$ has only one
solution $M \in (\delta/(\kappa + \beta), 1)$ which is given by
\[ M=\frac{(\delta-\delta_a)+\sqrt{\delta_a^2+\delta_a(\kappa+\beta -\delta)}}{\kappa+\beta}. \]
Observe that $M>\delta/(\kappa+\beta)$ if and only if $\sqrt{\delta_a^2 + \delta_a(\kappa+\beta-\delta)} > \delta_a$, which
is satisfied because we are assuming $\kappa+\beta > \delta$. Similarly, it can be seen that $M < 1$ if $\kappa+\beta > \delta$.
From (1)--(5) it follows that $a_1(s)$ has a global maximum in  $(\delta/(\kappa + \beta), 1)$ at $s=M$.
$\Box$

\begin{figure}
\centering
\includegraphics[scale=0.7]{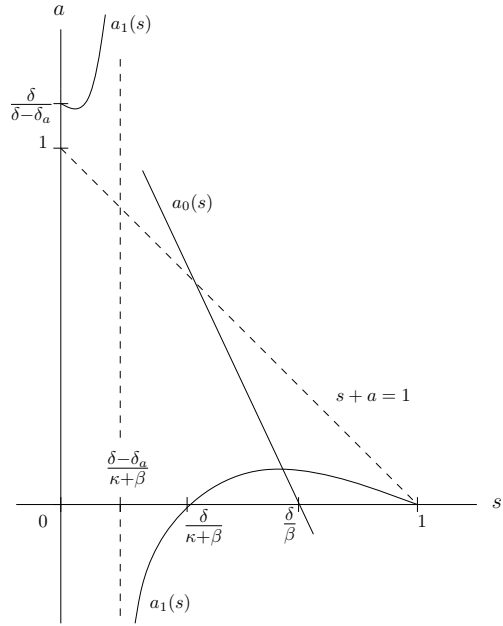}
\caption{Graphs of $a_0(s)$ and  $a_1(s)$ in the region $R$, when $\beta_a<\delta<\beta$ and $\delta > \delta_a$. The intersection of $a_0(s)$ with the boundary $s+a=1$ occurs at the point $(s^*_0, a^*_0)$ given by Theorem \ref{GB}. The non-existence of an endemic equilibrium  when $\delta > \beta$ is guaranteed because the slope of $a_0(s)$ is always less than $-1$ if $\beta > \beta_a$. 
\label{dibuix1}}
\end{figure}

In Figure~\ref{dibuix1} we can see a sketch of the curve $a_1(s)$ and the straight line $a_0(s)$ for $\delta_a < \delta < \beta$.
From Lemmas~\ref{rectapc} and \ref{propa1}, it immediately follows that $a_0(s)$ and $a_1(s)$ intersect each other exactly at one
point inside $R$ when $\delta<\beta$, which would correspond to the unique endemic equilibrium of system \eqref{edos} if it exists. So, 
by Lemma~\ref{rectapc}, we have:

\begin{corollary}\label{maxim1nedemic}
Assume $ \beta, \beta_a, \kappa, \delta, \delta_a > 0$ with $\beta > \delta$. Then the region $R$ contains at most one equilibrium point $(s^*, a^*)$ of \eqref{edos} different from DFE. If this equilibrium exists, its coordinates are given by
$$
(s^*, a^*)=\left( s^*, \frac{\delta-\beta s^*}{\beta_a} \right)
$$
with $s^*$ being the only solution of the equation $a_0(s)=a_1(s)$ with $s^* \in (\delta/(\kappa+\beta), 1)$.
\end{corollary}

To determine the existence of an endemic equilibrium we need to study the horizontal nullcline of system \eqref{edos}. The equation $da/dt=0$ leads to two possible curves:
\begin{equation}
a_2^\pm(s) := \frac{\kappa s+\delta_a+\beta_a(1-s) \pm \sqrt{(\kappa s - \beta_a (1-s))^2+2\delta_a (\kappa s + \beta_a (1-s))+\delta^2_a}}{2\beta_a}.
\label{a2}
\end{equation}
It is not difficult to see that $a_2^+(s)>1-s$ $ \forall s \in [0,1]$. Hence, $(s, a_2^+(s)) \notin R$ $ \forall s \in [0,1]$. So, from now on we will only consider
the nullcline $a_2(s):=a_2^-(s)$.

\begin{lemma}\label{propa2}
Assume $\beta_a, \beta, \kappa, \delta_a, \delta > 0$ with $\beta > \delta$. Then,
\begin{enumerate}
\item $0\le a_2(s)<1-s$ $\,\forall s \in [0, 1]$ with $a_2(0)=a_2(1)=0$.
\item $a_2'(0)>0$ and $-1<a_2'(1)<0$.
\item $a_2'(1)>a_1'(1)$.
\item $a_2(s)$ has a unique extremum in $(0,1)$, which is a maximum.
\end{enumerate}
\end{lemma}
\textit{Proof}. 
Statement 1 follows from direct computations. Let us prove statement 2. By implicit differentiation of Eq.~\eqref{a2} we get that
\begin{equation}\label{derivadaa2}
a_2'(s) = \frac{(\kappa-\beta_a)a_2-\kappa (1-2s)}{2 \beta_a a_2 - (\kappa s + \delta_a + \beta_a (1-s))},
\end{equation}
where the dependence of $a_2$ on $s$ has been omitted for simplicity of notation. From statement 1, we get that $a_2'(0)=\kappa/(\delta_a+\beta_a)>0$ and
$a_2'(1)=-\kappa/(\kappa+\delta_a)$, which lies between $-1$ and 0.

Let us prove statement 3. From Eq. \eqref{a1} we have that
$$
a_1'(1)=\frac{\kappa+\beta-\delta}{\delta-\delta_a-(\kappa+\beta)}.
$$
Then, $a_2'(1)>a_1'(1)$ if and only if $(\kappa+\delta_a+(\beta-\delta))\kappa<(\kappa+\delta_a)(\kappa+\beta-\delta)$, which is equivalent
to $(\beta-\delta)\kappa<(\kappa+\delta_a)(\beta-\delta)$. This inequality is obviously satisfied, since $\delta<\beta$ by hypothesis.

Finally, let us prove statement 4. It is enough to show that there is a unique value $s_0\in(0,1)$ such that $a_2'(s_0)=0$ since, in this case,
from statements 1 and 2 it follows that $s_0$ is a maximum of $a_2(s)$. From Eq. \eqref{derivadaa2}, the equation $a_2'(s)=0$ has as many solutions as
$$
a_2(s) = \frac{\kappa(1-2s)}{\kappa - \beta_a}
$$
with $a_2(s)$ given by Eq. \eqref{a2}. We can rewrite this equation as
\begin{eqnarray}
\left(\kappa - \beta_a + \frac{4 \beta_a \kappa}{\kappa - \beta_a}\right) s & + & \delta_a + \beta_a - \frac{2 \beta_a \kappa}{\kappa - \beta_a}
\nonumber \\
&=& \sqrt{(\kappa s - \beta_a (1-s))^2+2\delta_a (\kappa s + \beta_a (1-s))+\delta^2_a} \ . \label{zero}
\end{eqnarray}
The left-hand side of \eqref{zero} is linear in $s$, whereas the right-hand side is non-linear with at most one critical point for $s \in [0,1]$. Evaluating this equation at $s=0$ and $s=1$ we obtain:
\begin{eqnarray*}
\delta + \beta_a - \frac{2 \beta_a \kappa}{\kappa - \beta_a} < (>) \  \delta_a + \beta_a \quad if \quad \kappa > (<) \,\beta_a \quad \mbox{at} \quad s=0,\\
\kappa + \delta_a + \frac{2 \beta_a \kappa}{\kappa - \beta_a} > (<) \  \delta_a + \kappa \quad if \quad \kappa > (<) \,\beta_a  \quad \mbox{at} \quad s=1.
\end{eqnarray*}
Therefore, comparing the relative position of the end points of the curves defined by both sides of \eqref{zero} at $s=0$ and $s=1$, and taking into account their behaviour as functions of $s$, it follows the existence a unique intersection point $s_0$ between them in $(0,1)$.
$\Box$

We can summarise the results of this section in the following theorem: 
\begin{theorem}\label{E}
Assume $\beta_a, \beta, \kappa, \delta_a, \delta > 0$ with $\beta_a < \beta$. If $\delta<\beta$ then there exists a unique equilibrium of system \eqref{edos} in the interior of the region $R$, which is globally asymptotically stable. Otherwise ($\delta > \beta$), the DFE is the only equilibrium of \eqref{edos}
and is globally asymptotically stable.
\end{theorem}
\textit{Proof}.
From Lemmas~\ref{propa1} and \ref{propa2} it follows that, if $\beta > \delta$, $a_1(s)$ and $a_2(s)$ can intersect each other, at least, at one point in $s \in (0,1)$. Corollary~\ref{maxim1nedemic} tells us that they must intersect at exactly one point $(s^*, a^*)$ and characterizes its coordinates.

The global asymptotic stability of the interior equilibrium follows from the unstability of the DFE (the second eigenvalue of the Jacobian matrix $J(1,0)$ is $\lambda_2 = \beta - \delta > 0$), the invariance of the region $R \setminus L$, and Dulac's criterion for the nonexistence of closed orbits lying entirely in a simply connected region of $\mathbb{R}^2$ \cite{Perko}. In particular, if we consider the function $\varphi(s,a)=1/(1-s-a)$, which is continuously differentiable in $R \setminus L$, and denote by $f_1$ and $f_2$ the first and second component of the vector field defined by the rhs of \eqref{edos}, it follows that $\partial_s (\varphi  f_1) + \partial_a (\varphi  f_2) = -(\kappa + \beta + \beta_a) - \delta_a/(1-s-a) < 0$ for all $(s,a)$ in the interior of $R$. Therefore, Dulac's criterion guarantees the nonexistence of periodic orbits in the interior of $R$, and Poincar\'e-Bendixson theorem gives the global asymptotic stability of the interior equilibrium. The global asymptotic stability of the DFE when $\delta > \beta$ is given by Theorem~\ref{deltagebeta}.
$\Box$

\begin{figure}
\begin{center}
\begin{tabular}{cc}
\hspace{0cm}
	\includegraphics[scale=0.77]{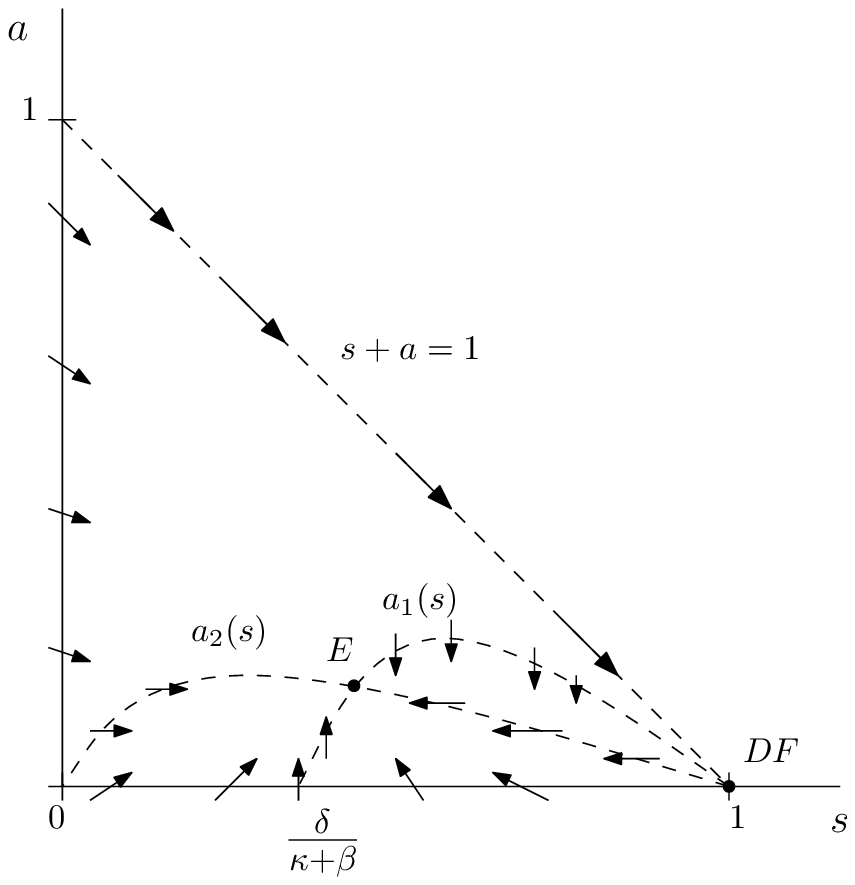}
&
	\includegraphics[scale=0.35]{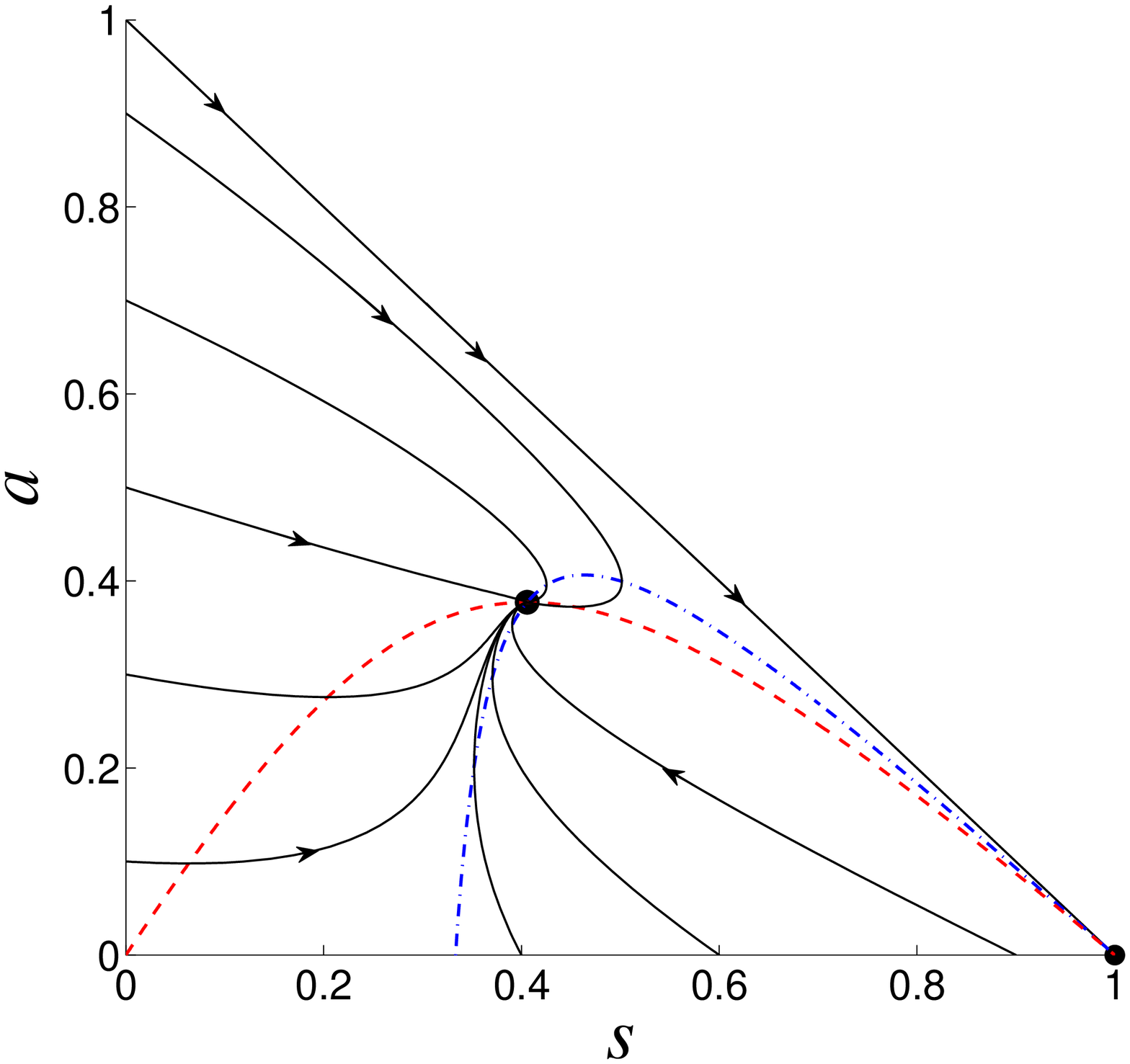}
\end{tabular}
\caption{Left: Sketch of the vector field associated to system \eqref{edos} for $\delta<\beta$. The endemic equilibrium $E$ is globally asymptotically stable 
whereas the disease-free equilibrium $DF$ is a saddle point. Right: Phase portrait of system \eqref{edos} for $\delta=4$, $\delta_a = 0.5$, $\beta=8$, $\beta_a=2$, and $\kappa=4$. Red dashed line is the graph of $a_2(s)$ and blue dash-dotted line is that of $a_1(s)$. 
\label{dibuix2}}
\end{center}
\end{figure}

To finish the sketch of the phase portrait of system \eqref{edos}, we compare the slopes of the nullclines at $(s,a)=(1,0)$ and that of the
eigenvector $v_2=(v_2^1, v_2^2)$ of the Jacobian matrix $J(1,0)$ associated to $\lambda_2=\beta-\delta$ (see Theorem \ref{deltagebeta}).
When $\delta<\beta$ we have that
\[ \frac{v_2^2}{v_2^1} = \frac{\kappa}{\delta-\delta_a-(\kappa+\beta)} > \frac{\kappa}{-\kappa-\delta_a} = a_2'(1)
> \frac{\kappa+\beta-\delta}{\delta-\delta_a-(\kappa+\beta)} = a_1'(1), \]
where the last inequality has been stated in the proof Lemma~\ref{propa2}. Conversely, when $\delta>\beta$ the previous inequalities are fulfilled in the
opposite direction. When $\delta=\beta$, the endemic equilibrium $E$ bifurcates from the DFE  because, at this point, $\lambda_2=0$ and the slopes of the nullclines at $s=1$ are equal to each other (and equal to the one of the eigenvector $v_2$). The left panel of Figure~\ref{dibuix2} depicts a sketch of the vector field associated to system \eqref{edos} for $\delta< \beta$. The right panel shows the phase portrait of this system for a particular choice of the parameters values that leads to a similar relative position of the nullclines. 

\begin{figure}[ht]
\centering
\includegraphics[scale=0.45]{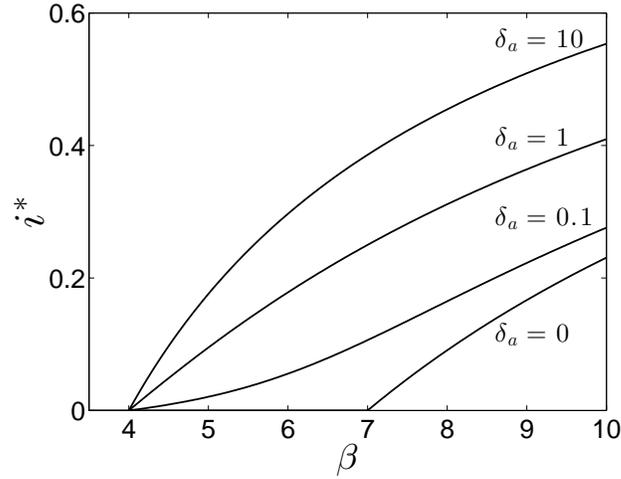}
\caption[Phase]{Behaviour of the fraction of infectious nodes $i^*$ at equilibrium as a function of the transmission rate $\beta$ for different values of $\delta_a$.  For $\delta_a=0$, bifurcation occurs at the second epidemic threshold, here given by $\beta=7$. For $\delta_a > 0$, $i^*(\beta) > 0$ for all $\beta > \delta$. The other parameters are: $\delta=4$, $\beta_a=2$, and $\kappa=3$.}
\label{i*-i(t)}
\end{figure}

The previous inequalities also show that, for $\kappa + \beta > \delta$, the slopes of both nullclines at $s=1$ tend to $-1$ as $\delta_a \to 0$ and, hence, the two curves tend to be very close to each other and to the boundary $s+a=1$ for values of $s$ close to 1. Moreover, as $a'_1\left(\frac{\delta}{\kappa+\beta}\right) \to \infty$ when $\delta_a \to 0$ (the location of the vertical asymptote of the graph of $a_1(s)$ tends to $\delta/(\kappa+\beta)$ when $\delta_a \to 0$, see Figure \ref{dibuix1}), and the nullclines intersect each other at a point $(s^*, a^*)$ on the graph of $a_0(s)$, which does not depend on $\delta_a$, it follows that $(s^*, a^*)$ moves along the straight line $(s, a_0(s))$ towards the boundary $s+a=1$ as $\delta_a$ decreases. When $\beta_a < \delta < \beta$, the intersection of $a_0(s)$ with the boundary $s+a=1$ occurs at $\left(\frac{\delta - \beta_a}{\beta-\beta_a}, \frac{\beta - \delta}{\beta-\beta_a} \right)$, i.e., at the point $(s^*_0, 1-s_0^*)$ that splits this boundary into two regions when $\delta_a =0$ (cf. Theorem \ref{GB}). If, in addition, $\kappa > \kappa^*$ then $(s^*,a^*)$ approaches $(s^*_0, 1-s_0^*)$ as $\delta_a \to 0$ because this condition is equivalent to $s^*_0 > \delta/(\kappa + \beta)$. Therefore,  the fraction of infectious nodes at equilibrium $i^*=1-s^*-a^*$ tends to 0 as $\delta_a \to 0$ whenever $\beta_a < \delta < \beta$ and $\kappa > \kappa^*$.  On the other hand, using again the dependence of $a'_1(1)$ and $a'_2(1)$ on $\delta_a$ and the fact that $(s^*, a^*)$ belongs to the graph of $a_0(s)$, it follows that $i^* \to 1-\delta/\beta$ as $\delta_a \to \infty$, the endemic equilibrium of an SIS model when $\beta > \delta$. Figure  \ref{i*-i(t)} shows the dependence of the fraction $i^*$ of infected nodes at equilibrium on $\beta$ for different values of $\delta_a$. 

\begin{figure}[ht]
\begin{center}
\begin{tabular}{cc}
\hspace{-1cm}
	\includegraphics[scale=0.35]{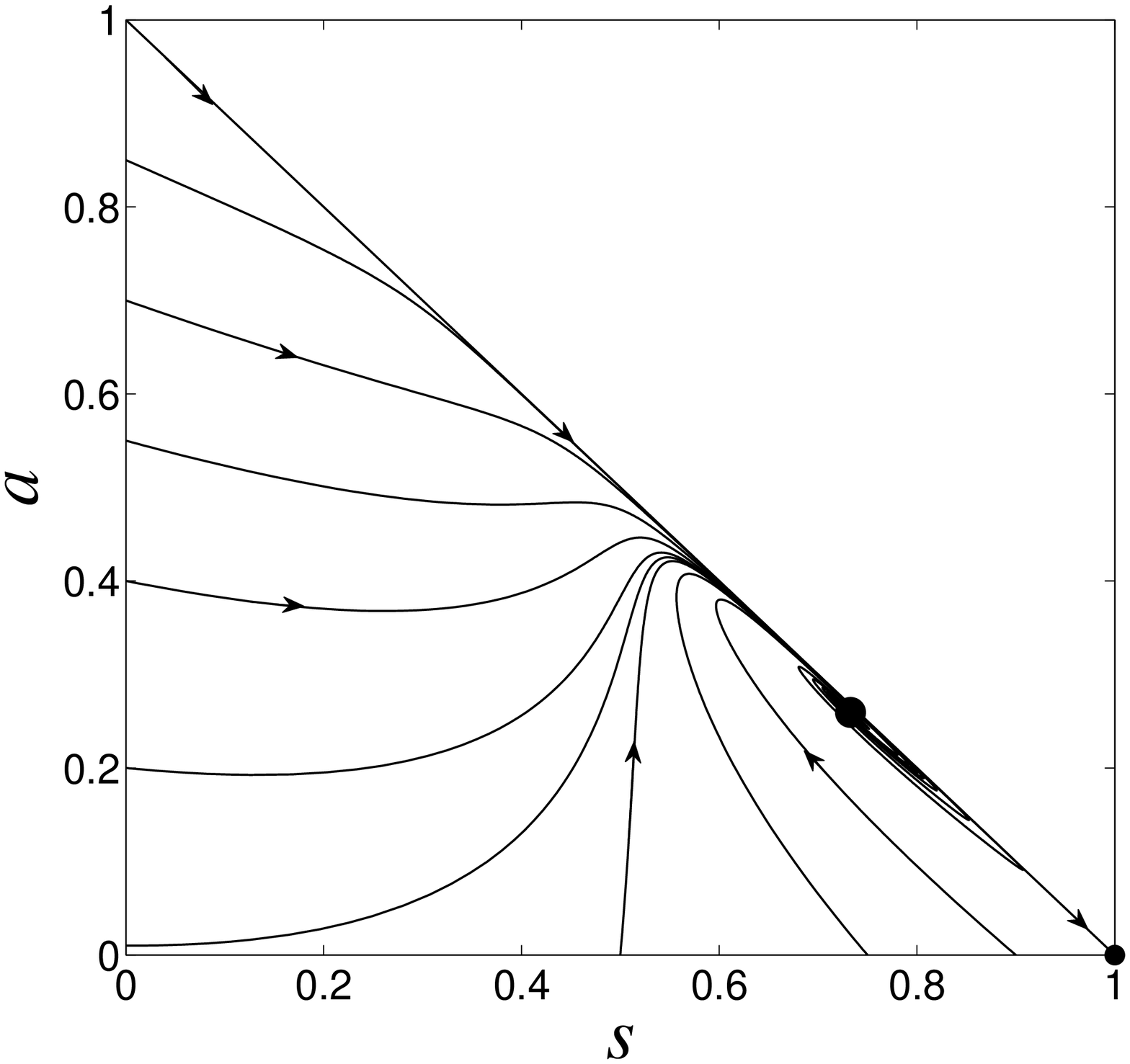}
&
	\includegraphics[scale=0.35]{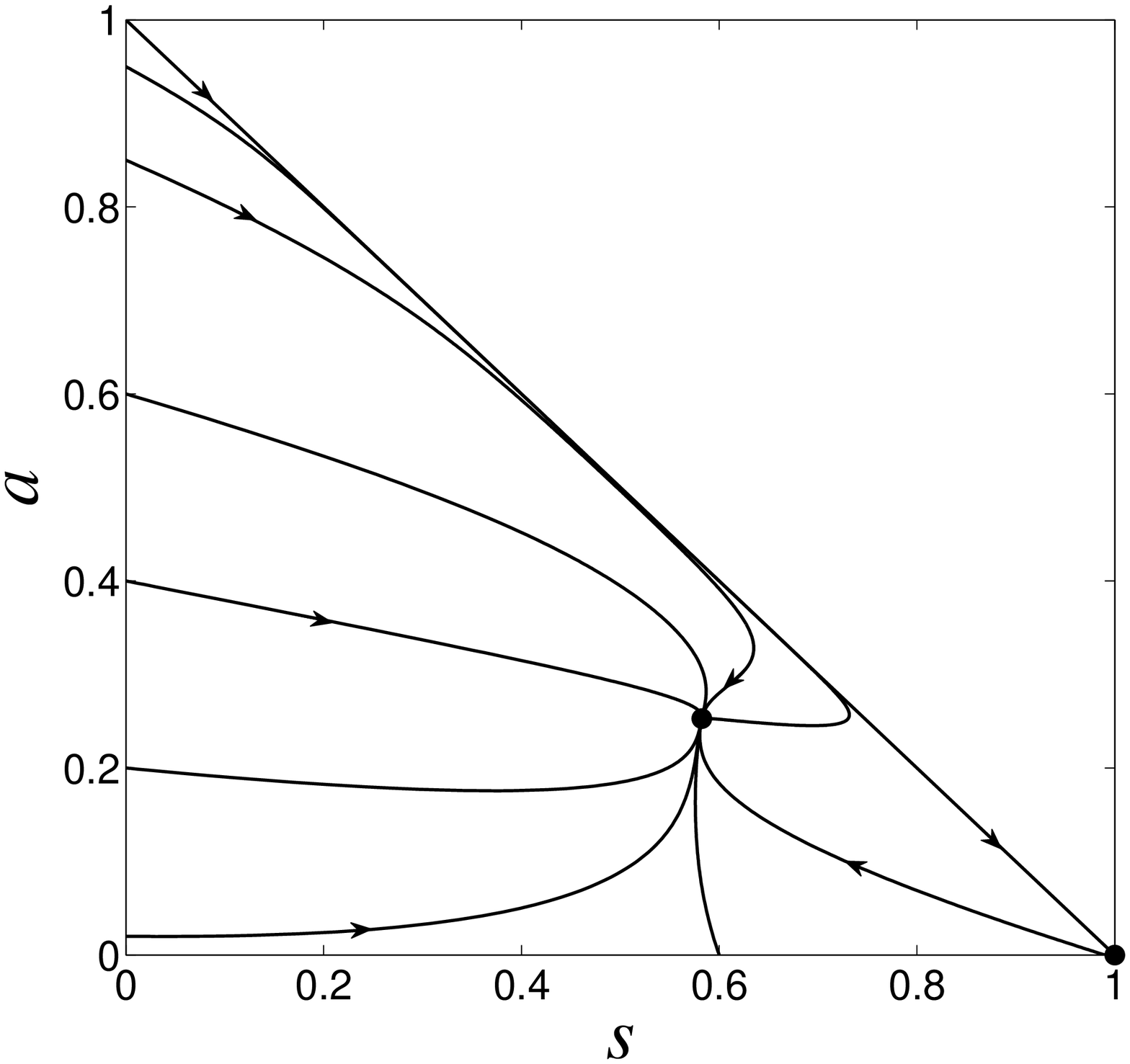}
\end{tabular}
\caption[Phase]{Phase portrait of system \eqref{edos} for a very small rate of awareness decay and the same parameters values as those used in panels of Figure \ref{fig:Case2}. In both panels, the endemic equilibrium attracts every trajectory inside $R$ and the disease-free equilibrium $(1,0)$ is a saddle point. Left: Trajectories corresponding to minor outbreaks in left panel of Figure \ref{fig:Case2} now spiral in towards an endemic equilibrium which is very close to the boundary $s+a=1$, i.e., an equilibrium with a very low fraction of infectious nodes: $i^*=0.0077$. As $\delta_a \to 0$, this endemic equilibrium approaches the point $(s^*_0, 1-s^*_0)$ on the boundary in Figure \ref{fig:Case2}. Right: In contrast to what happens in right panel of Figure \ref{fig:Case2}, the basin of attraction of the endemic equilibrium $(s^*, a^*)=(0.5824, 0.2528)$ is the whole interior of $R$. Parameters: $\delta=4$, $\beta_a=2$, $\delta_a=0.05$, and $\beta=4.75$, $\kappa=3$ (left) and $\beta=6$, $\kappa=1$ (right).} 
\label{Comparison-fig:Case2}
\end{center}
\end{figure}

We have already observed in Figures~\ref{fig:Case2} and \ref{fig:evolution_of_cases} that the way trajectories approach an equilibrium can be sensitive to parameters values and to initial conditions when $\delta_a=0$. Figure~\ref{Comparison-fig:Case2} shows the phase portaits of system \eqref{edos} for the same parameters values as in Figure~\ref{fig:Case2} and $\delta_a=0.05$.  They clearly show that the solutions tend to an endemic equilibrium and, so, that the two epidemic models have different qualitative behaviours, as expected from the structural instability of system~\eqref{edos-nodecay} (see Peixoto's theorem in \cite{Perko}). For instance, trajectories corresponding to minor outbreaks in Figure~\ref{fig:Case2} are now replaced by damped oscillations converging to an endemic equilibrium which is close to the boundary $s+a=1$ and attracts every trajectory inside the region $R$ (left panel). Similarly, trajectories with a high initial fraction of aware nodes that ended up at a disease-free equilibrium when $\delta_a=0$, now tend to an endemic equilibrium with $i^*=0.165$ which is globally stable (right panel).  

\begin{figure}[ht]
\begin{tabular}{ll}
\hspace{-1cm}
	\includegraphics[scale=0.4]{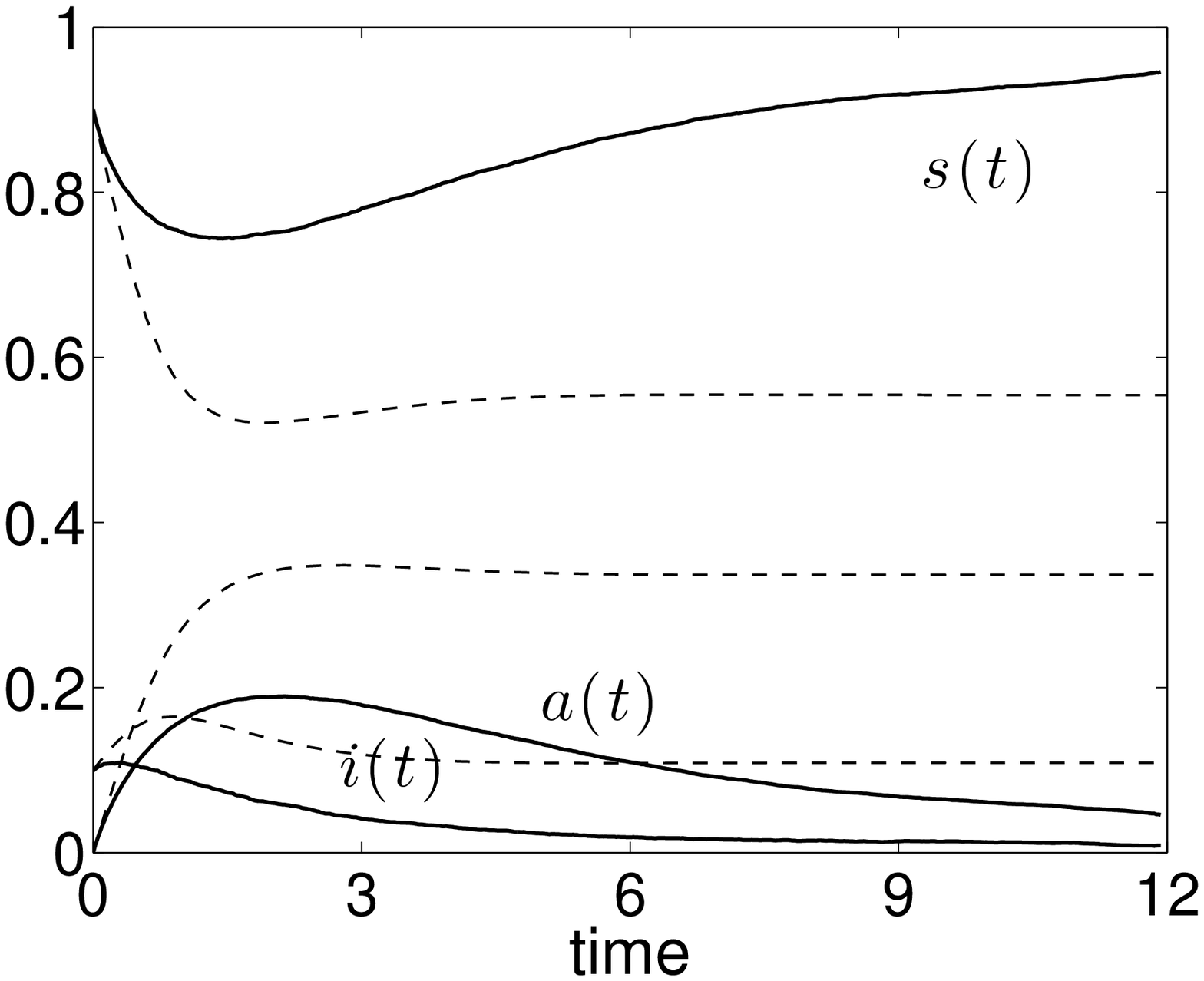}
&
	\includegraphics[scale=0.4]{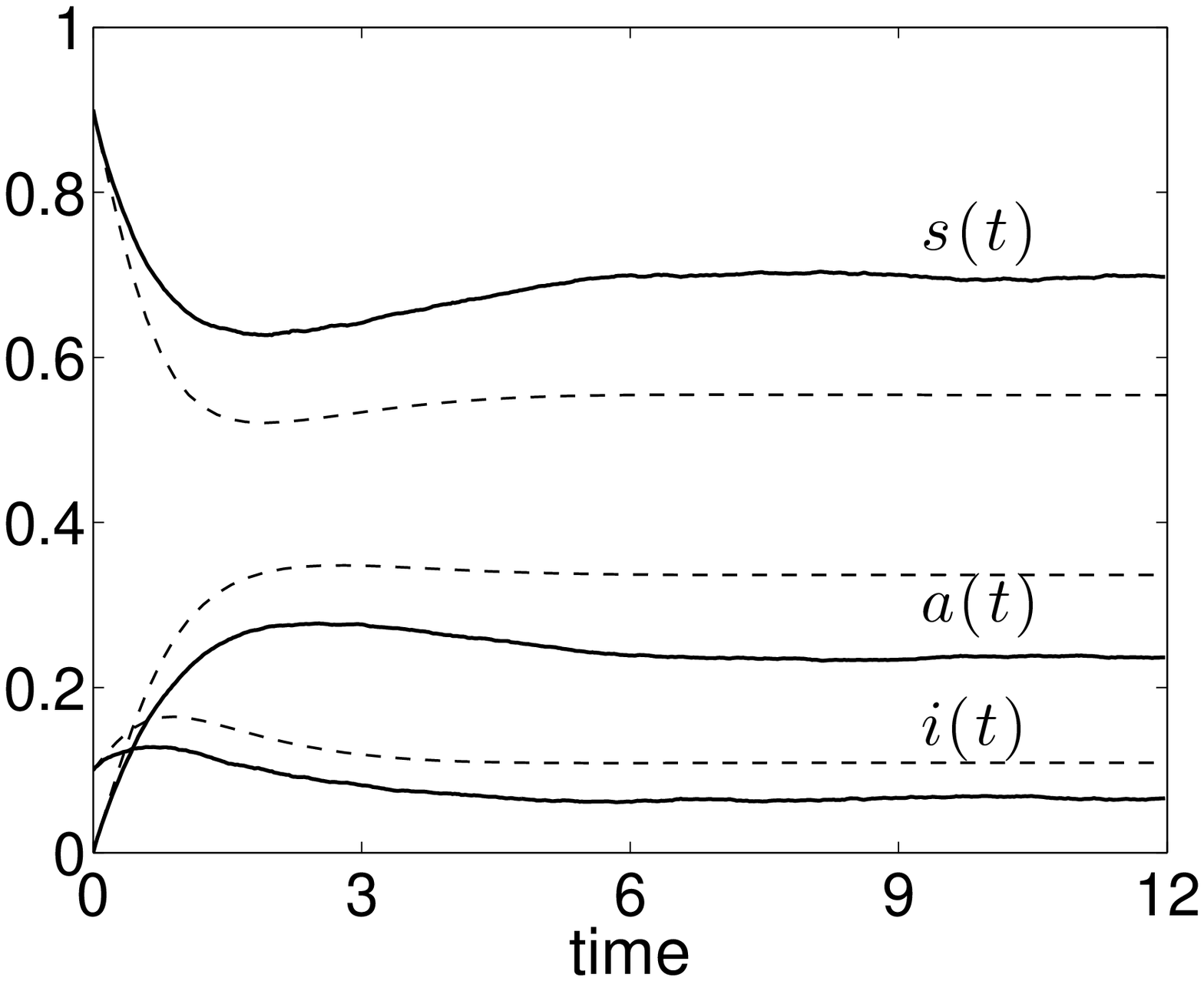}
\\
\hspace{-1cm}
	\includegraphics[scale=0.4]{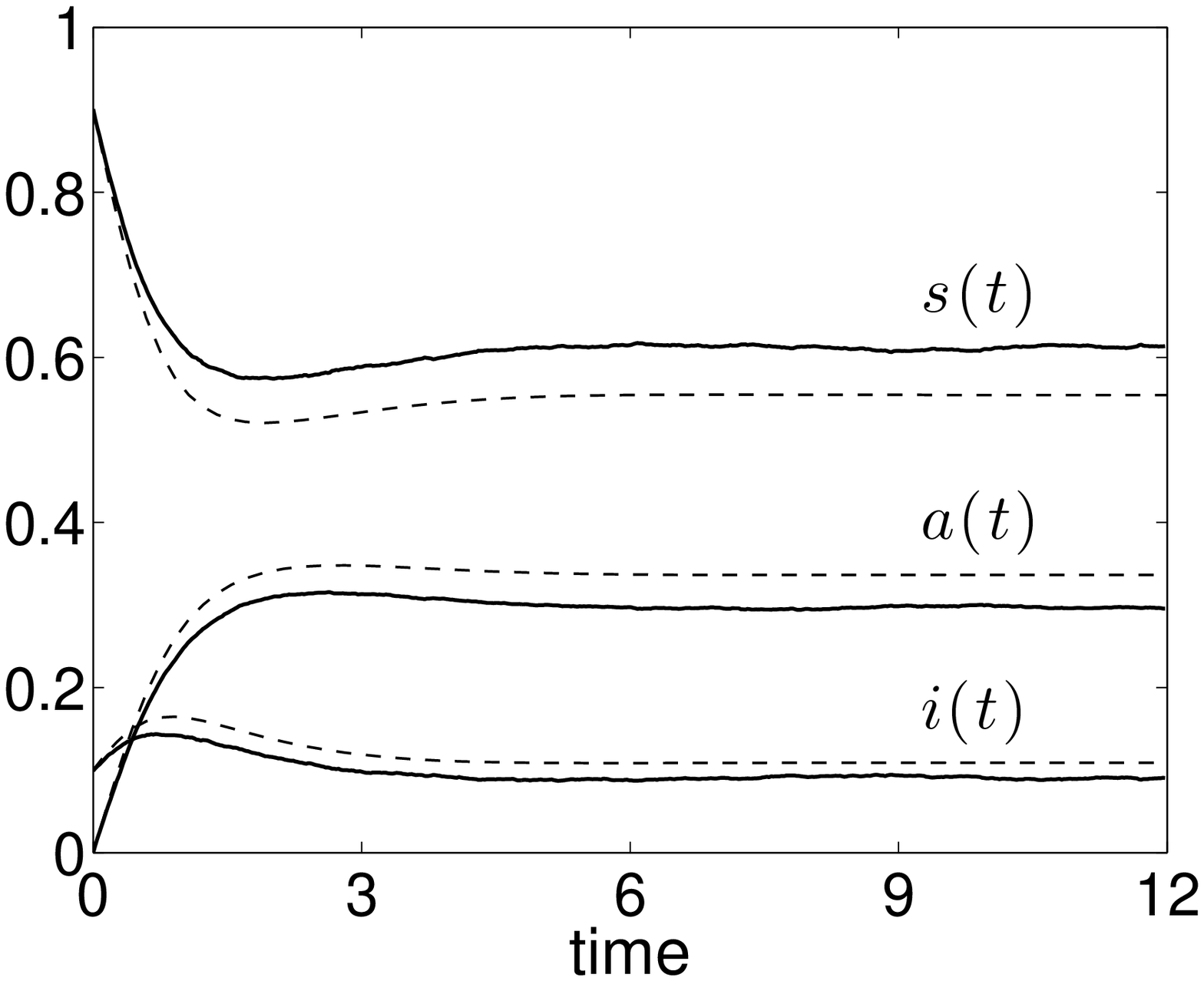}
&
	\includegraphics[scale=0.4]{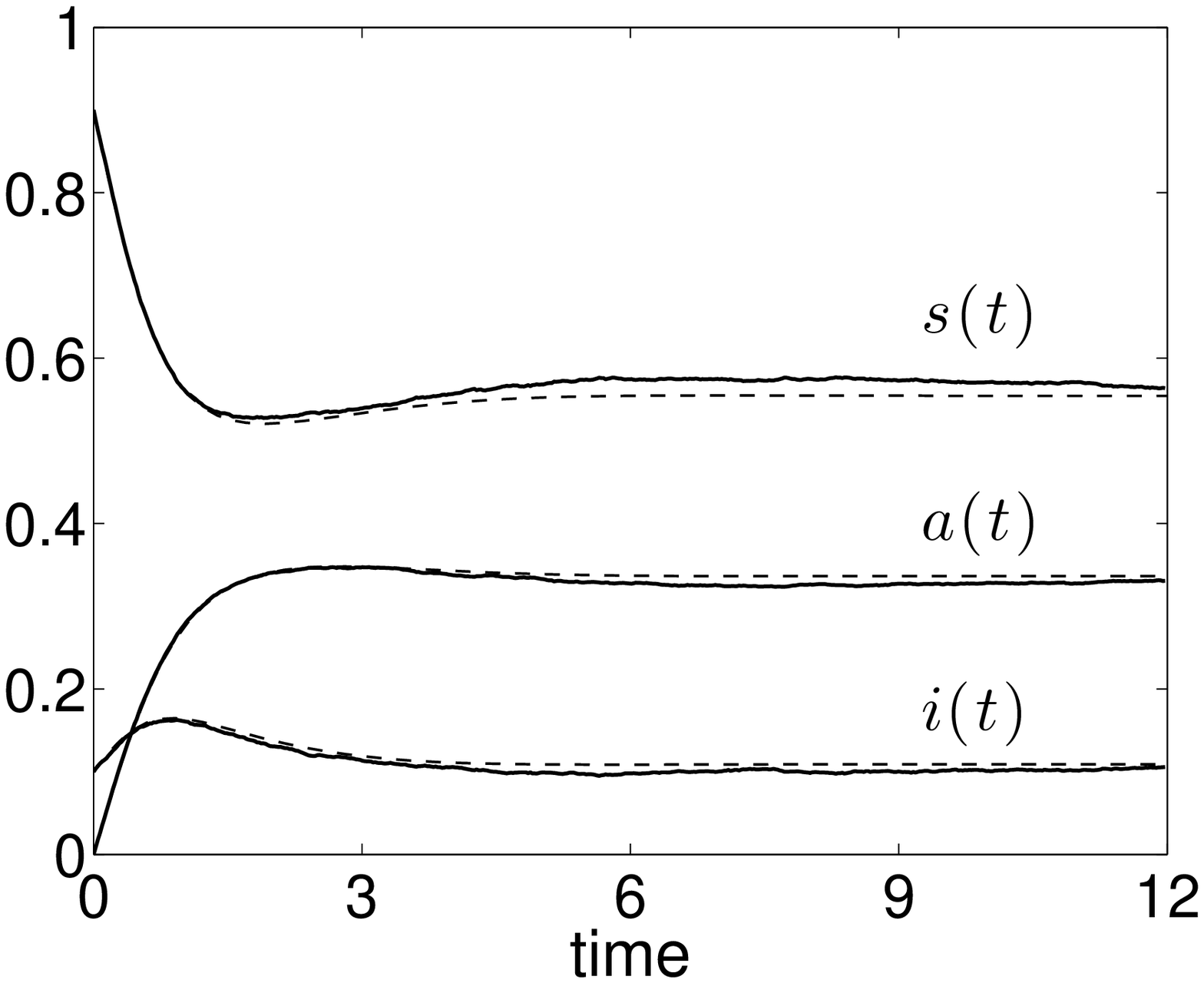}
\end{tabular}
\caption{
Evolution of the fraction of infectious ($i$), susceptible ($s$), and aware ($a$) individuals with awareness decay for the same rates values 
as in Figure \ref{fig:comparison}, i .e., when a minor outbreak is predicted for $\delta_a=0$.
Dashed lines: solutions to system~\eqref{edos} with initial condition $s(0)=0.90$, $i(0)=0.1$ and $a(0)=0$.
Solid lines: stochastic simulations over regular random networks of size 1000 and degree 5 (top left), 10 (top right), 20 (bottom left), and 
fully connected (bottom right). Simulation outputs averaged over 100 runs with 10\% of randomly infected individuals and 90\% of susceptible ones at $t=0$. Parameters: $\delta=4$, $\delta_a = 0.5$, $\beta=6$, $\beta_a=2$, $\kappa=4$.
}.
\label{fig:comparison2}
\end{figure}

In Figure~\ref{fig:comparison2} we check the accuracy of the model when an endemic equilibrium is predicted. As in Figure \ref{fig:comparison}, we show the time evolution of the fraction of infectious, aware and susceptible individuals, averaged over 100 realizations of continuous-time stochastic simulations of an epidemic performed on regular random networks of size 1000. In each realization, 100 susceptible individuals are initially infected uniformly at random. The parameter values are the same as in that figure except for $\delta_a$ that now is positive and equal to 0.5. This value implies an average duration of the awareness period eight times longer than the infectious period. Remarkably, the endemic equilibrium is not observed in networks of very low degree ($k=5$ in the top-left panel) because of the low force of infection during the early stage of an epidemic. Note that the presence of infectious individuals around those initially infected reduces significantly the potential transmission of the infection when the nodal degree is low. This saturation in the transmission is clearly less marked for $k=10$ and disappears in fully connected networks, for which the agreement between simulations and the model is almost perfect. For degree values about 20 the qualitative behaviour observed in the simulations is quite in agreement with the model.


\section{Discussion and conclusions}

In this paper we have considered the susceptible-aware-infectious model proposed in \cite{Sahneh11,Sahneh12a} on regular random networks. For this type of networks, we derived a simple mean-field model and proved that it has the same solutions than the original node-based model when initial infections of susceptible individuals occur uniformly at random. This exact correspondence refutes previous claims about the underestimation of the rate of infection by mean-field versions of node-based SIS models defined on regular networks \cite{Mieghem}. In fact, numerical simulations showed that, when initial conditions are given by clustered infections, i.e., infections are not uniformly at random, the mean-field SAIS model overestimates the initial epidemic growth predicted by its node-based counterpart (see Figure \ref{MF-vs-Intertwined}). On the other hand, the agreement of model predictions with the stochastic simulations on regular random networks increases with the degree of the network, and it becomes almost perfect in fully connected networks, as expected. It is well-known that mean-field models assume homogeneous mixing of individuals and, hence, overestimate the number of susceptible nodes around the infectious ones when networks are not fully connected. In the SAIS model, such an overestimation affects the predicted number of both infectious and aware individuals. 

In this mean-field model, the segment $L=\{(s,a) \in \mathbb{R}^2 \ | \ s+a=1, 0 \le s \le 1\}$ defines a continuum of (disease-free) equilibria. When $\beta < \delta$, any initial number of infectious individuals will tend monotonously to zero and, so, no minor outbreaks are possible. This is the typical situation of an epidemic extinction and corresponds to what has been called ``quick die out" in \cite{Sahneh12a}. For $\beta > \delta > \beta_a$, the equilibria on the segment $L$ close to $(1,0)$ become unstable. Interestingly, this happens before the appearance of an endemic equilibrium which bifurcates from the point $(s^*_0,1-s^*_0) \in L$ such that $s^*_0 = (\delta-\beta_a)/(\beta-\beta_a)$. Taking the awareness rate $\kappa$ as a tuning parameter, the condition for the bifurcation to occur defines a second epidemic threshold given by $\kappa^* = \beta_a(\beta - \delta)/(\delta-\beta_a)$. This threshold is the same that the one given by the expression (6) in \cite{Sahneh12a} for system \eqref{markov} with $\delta_a = 0$ if one realizes that the dominant eigenvalue $\lambda_1$ of the adjacency matrix equals the nodal degree $k$ for regular random networks. So, for $\beta > \delta > \beta_a$, $\kappa \ge \kappa^*$, and assuming a small fraction of initially infectious and aware individuals, any trajectory tends to an equilibrium on $L$ after an initial increase in the number of infectious individuals (see left panel in Figure \ref{fig:Case2}). This scenario has been defined as "slow die-out" of the epidemic in \cite{Sahneh12a} and leads to a final population with a significant number of aware individuals. For $\beta > \delta > \beta_a$ and $\kappa < \kappa^*$, the awareness rate is not high enough and the system has an endemic equilibrium attracting all trajectories nearby. However, this equilibrium is not globally stable because the infection rate of aware individuals $\beta_a$ is low enough to prevent the occurrence of an endemic equilibrium in a population mostly consisting of aware individuals (see right panel of Figure 2). Finally, for $\beta > \beta_a > \delta$, the susceptibility of aware individuals is high enough to allow for an endemic equilibrium attracting all trajectories with a positive initial fraction of infectious individuals.

This information about the behaviour of the mean-field model \eqref{edos-nodecay} is, in fact, encapsulated in the expression \eqref{solution} of their trajectories. From it one can obtain a description of their transient behaviour. For instance, from this expression it is clear that trajectories in the phase plane never oscillate, in contrast to what happens to the solutions of system \eqref{edos} for some combinations of the parameters. It also offers an analytical expression of the minor outbreaks occurring between the first and second epidemic threshold.  Moreover, when the system is above the second threshold and the endemic equilibrium only attracts nearby trajectories, we can also have a precise determination of its the basin of attraction.    

On the other hand, it is reasonable to expect that people forget their awareness ($\delta_a > 0$), especially with a low prevalence of the disease \cite{Wei}.
So, we extended the mean-field model \eqref{edos-nodecay} to include awareness decay at a constant rate $\delta_a$.  From a mathematical point of view, the resulting system \eqref{edos} is a (continuously differentiable) perturbation of system \eqref{edos-nodecay}, which turns out to be structurally unstable because of its continuum of (disease-free) equilibria (see Peixoto's theorem in \cite{Perko} for a full characterization of structurally stable planar systems defined on compact sets of $\mathbb{R}^2$). The main consequence of this perturbation is that the epidemic dynamics now do not have a second threshold, and the global behaviour of solutions reduces to the standard one in many epidemic models: a unique disease-free equilibrium (with no aware individuals) which is globally asymptotically stable (GAE) for $\beta < \delta$, while it becomes unstable and it appears a GAE endemic equilibrium for $\beta > \delta$. 

Similar changes in the dynamics are also observed when one compares classic epidemic models (with no demography) in which the susceptible class can be replenished by processes like recovery or loss of immunity (for instance, SIS, SIRS, and SEIRS models) with those in which it is not (SIR and SEIR models) \cite{EK}. From the point of view of dynamical systems, the latter are also structurally unstable because of their continuum of disease-free equilibria, and it is well known that smooth perturbations of a structural unstable system, no matter how small, can modify the phase portrait of the unperturbed system. However, in the SAIS model, the susceptible class can be partially  renewed via recovery of infectious individuals, even without awareness decay. This is the reason why, in contrast to the SIR model, disease can persist if $\delta_a=0$ as long as the awareness rate $\kappa$ is low enough or, alternatively, susceptibility of aware individuals is high enough. 

What are the consequences of this change in the dynamics of the SAIS model on the prevalence of the disease? For  $\delta < \beta_a < \beta$, the epidemic dynamics is governed by the presence of a GAE endemic equilibrium in both models and, so, there are no remarkable qualitative differences with respect to the prevalence of the disease. However, for $\beta_a < \delta < \beta$ and small values of $\delta_a$ two interesting changes are noticeable (see Figure \ref{Comparison-fig:Case2}). Below the second epidemic threshold of system \eqref{edos-nodecay} (i.e., for $\delta < \beta < \delta + \kappa (\delta - \beta_a) / \beta_a$), solutions $(s(t), a(t))$ of system \eqref{edos} spiral in towards the endemic equilibrium, which is located near the boundary $s+a=1$, instead of representing minor outbreaks occurring when the initial fraction of aware individuals is negligible and the inital fraction of infectious individuals is very small (see Figure \ref{fig:Case2}). Above the second threshold ($\beta > \delta + \kappa (\delta - \beta_a) / \beta_a$), the main difference between solutions arises for trajectories starting with a high fraction of aware individuals and a very small of susceptibles ones. According to system \eqref{edos-nodecay}, these trajectories tend to a disease-free equilibrium whereas, in system \eqref{edos}, they tend to an endemic equilibrium with a significant  fraction of infectious individuals when $\beta \gg \delta$. The interesting fact is that, for very small $\delta_a$, such trajectories remain close to the boundary $s+a=1$ for a long time during which the epidemic seems to be eradicated from de population, before they eventually approach the endemic equilibrium. Therefore, regarding the prevalence of the disease, these solutions have a transient behaviour similar to that of the solutions of system \eqref{edos-nodecay} with the same initial conditions, but they have a completely different asymptotic behaviour. 

In this paper, we have assumed a constant rate of awareness decay. One could think of this hypothesis to be quite restrictive, and consider more general dependences of this rate on the disease prevalence. However, as long as they constitute small enough smooth perturbations of system \eqref{edos}, a qualitatively similar phase portraits will result because this system is structurally stable. The addition of a term accounting for the creation of new aware individuals by already aware individuals is another possible extension of the original model. Such a term has been considered in previous papers dealing with epidemic models and information transmission \cite{Funk10, Granell, Kiss10} and allows for a change of the classic epidemic threshold as long as aware individuals are able to self-sustain their numbers in the absence of disease, that is, when awareness behaves as second epidemic spreading across the population. 

More sophisticated models consider networks with diverse relationships (layers) among their nodes \cite{Cardillo, Granell, Saumell, Wang13}. These interconnected networks are used, for instance, to model the transmission of multiple pathogens on the same population \cite{Funk10b}, or the simultaneous spread of an infectious agent and information about the health state of individuals. The analysis of these elaborated network models have shown that new results emerge from the interaction of the interconnected networks \cite{Saumell}. However, such an added complexity can hide some aspects of the dynamics that are rooted in the basic ingredients of the transmission process itself. In \cite{Sahneh14, Sahneh12b}, a layer for information dissemination was introduced in the epidemic model, in addition to the one of physical contacts among individuals. As in \cite{Sahneh11, Sahneh12a} no awareness decay was assumed in these works, and the existence of a second epidemic threshold related to the preventive behaviour of aware individuals was proven. This second threshold is certainly inherited from the one-layer version of the model, and it is very likely that other aspects of the dynamics are inherited as well. Our results strongly suggest that the introduction of an awareness decay into the two-layer version of the model would have the same implications than for the one-layer model. In particular, it would result in the disappearance of the second epidemic threshold.
 
\section*{Acknowledgments}
This work has been partially supported by the research grant MTM2011-27739-C04-03 of the Spanish government (D.J., J.S.), the project 2009-SGR-345 (J.S.) of the Generalitat de Catalunya,  and IMA Collaborative Grant (SGS01/13), UK,  (I.K., J.S.).

\pagenumbering{gobble}

\newpage

\newpage

\newpage

\newpage

\newpage

\newpage

\newpage

\newpage

\newpage

\end{document}